%% file: TOI1710b_corr.tex
\documentclass[longauth]{aa} 

\usepackage{graphicx}
\usepackage{txfonts}
\usepackage[colorlinks=true,citecolor=blue]{hyperref}
\DeclareUnicodeCharacter{2212}{-}

\begin{document}

   \title{Revisiting the warm sub-Saturn TOI-1710b}

   \subtitle{The impact of stellar activity on the mass measurement}

   \author{J.~Orell-Miquel\inst{\ref{ins:iac},\ref{ins:ull}}
          \and
          I.~Carleo\inst{\ref{ins:iac},\ref{ins:ull}}
          \and
          F.~Murgas\inst{\ref{ins:iac},\ref{ins:ull}}
          \and
          G.~Nowak\inst{\ref{ins:ia_ncu},\ref{ins:iac},\ref{ins:ull}}
          \and
          E.~Pall\'e\inst{\ref{ins:iac},\ref{ins:ull}}
          \and
          R.~Luque\inst{\ref{ins:iaa}}
          \and
          T.~Masseron\inst{\ref{ins:iac},\ref{ins:ull}}
          \and
          J.~Sanz-Forcada\inst{\ref{ins:astro}}
          \and
          D.~Dragomir\inst{\ref{ins:new_mex}}
          \and
          P. A.~Dalba\inst{\ref{ins:paul1},\ref{ins:paul2}}
          \and
          R.~Tronsgaard\inst{\ref{ins:denmark}}
          \and
          J.~Wittrock\inst{\ref{ins:mason2}}
          \and
          K.~Kim\inst{\ref{ins:mason1}}
          \and
          C.~Stibbards\inst{\ref{ins:mason2}}
          \and
          K. I.~Collins\inst{\ref{ins:mason2}}
          \and
          P.~Plavchan\inst{\ref{ins:mason2}}
          \and
          S. B.~Howell\inst{\ref{ins:nasa2}}
          \and
          E.~Furlan \inst{\ref{ins:nasa1}}
          \and
          L. A.~Buchhave\inst{\ref{ins:denmark}}
          \and
          C. L.~Gnilka \inst{\ref{ins:nasa2}}
          \and
          A. F.~Gupta\inst{\ref{ins:arv1},\ref{ins:arv2}}
          \and
          Th.~Henning\inst{\ref{ins:Max}}
          \and
          K. V.~Lester\inst{\ref{ins:nasa2}}
          \and
          J. E.~Rodriguez\inst{\ref{ins:Michigan}}
          \and
          N. J.~Scott \inst{\ref{ins:nasa2}}
          \and
          H. P.~Osborn\inst{\ref{ins:nccr},\ref{ins:tess1}}
          \and
          S.~Villanueva Jr.\inst{\ref{ins:tess1},\ref{ins:pappalardo}}
          \and
          S.~Seager\inst{\ref{ins:tess1},\ref{ins:sara1},\ref{ins:sara2}}
          \and
          J.~N.\, Winn\inst{\ref{ins:tess2}}
          \and
          J. M.~Jenkins\inst{\ref{ins:nasa2}}
          \and
          R.~Vanderspek\inst{\ref{ins:tess1}}
          \and
          D. W.~Latham\inst{\ref{ins:smith}}
          \and
          P.~Rowden\inst{\ref{ins:royal}}
          \and
          D.~Watanabe\inst{\ref{ins:fred}}
          \and
          G.~Torres\inst{\ref{ins:smith}}
          \and
          C. J.~Burke\inst{\ref{ins:tess1}}
          \and
          T.~Daylan\inst{\ref{ins:daylan}}
          \and
          T.~Barclay\inst{\ref{ins:tess4}}
          \and
          J. D.~Twicken\inst{\ref{ins:seti},\ref{ins:nasa2}}
          \and
          G. R.~Ricker\inst{\ref{ins:tess1}}
          }

   \institute{
        \label{ins:iac} Instituto de Astrofísica de Canarias (IAC), 38205 La Laguna, Tenerife, Spain\\
        \email{jom@iac.es}
        \and
        \label{ins:ull} Departamento de Astrofísica, Universidad de La Laguna (ULL), 38206 La Laguna, Tenerife, Spain
        \and
        \label{ins:ia_ncu} Institute of Astronomy, Faculty of Physics, Astronomy and Informatics, Nicolaus Copernicus University, Grudzi\c{a}dzka 5, 87-100 Toru\'n, Poland
        \and
        \label{ins:iaa} Instituto de Astrof\'isica de Andaluc\'ia (IAA-CSIC), Glorieta de la Astronom\'ia s/n, 18008 Granada, Spain
        \and
        \label{ins:denmark} DTU Space, National Space Institute, Technical University of Denmark, Elektrovej 328, DK-2800 Kgs. Lyngby, Denmark
        \and
        \label{ins:mason1} Thomas Jefferson High School, 6560 Braddock Rd, Alexandria, VA 22312 USA
        \and
        \label{ins:mason2} George Mason University, Department of Physics and Astronomy, 4400 University Drive MS 3F3, Fairfax, VA 22030 USA
        \and
        \label{ins:astro} Centro de Astrobiolog\'{i}a (CSIC-INTA), ESAC Campus, Camino Bajo del Castillo s/n, Villanueva de la Ca\~{n}ada, E-28692 Madrid, Spain
        \and
        \label{ins:nasa1} NASA Exoplanet Science Institute, Caltech/IPAC, Mail Code 100-22, 1200 E. California Blvd., Pasadena, CA 91125, USA
        \and
        \label{ins:nasa2} NASA Ames Research Center, Moffett Field, CA 94035, USA
        \and
        \label{ins:Michigan} Department of Physics and Astronomy, Michigan State University, East Lansing, MI 48824, USA
        \and
        \label{ins:Max} Max Planck Institute for Astronomy, Königstuhl 17, D-69117 Heidelberg, Germany
        \and
        \label{ins:paul1} Department of Earth and Planetary Sciences, University of California Riverside, 900 University Ave, Riverside, CA 92521, USA
        \and
        \label{ins:paul2} NSF Astronomy and Astrophysics Postdoctoral Fellow
        \and
        \label{ins:arv1} Department of Astronomy \& Astrophysics, 525 Davey Laboratory, The Pennsylvania State University, University Park, PA, 16802, USA
        \and
        \label{ins:arv2} Center for Exoplanets and Habitable Worlds, 525 Davey Laboratory, The Pennsylvania State University, University Park, PA, 16802, USA
        \and
        \label{ins:new_mex} Department of Physics and Astronomy, University of New Mexico, 210 Yale Blvd NE, Albuquerque, NM 87106, USA
        \and
        \label{ins:sara1} Department of Earth, Atmospheric and Planetary Sciences, Massachusetts Institute of Technology, Cambridge, MA 02139, USA
        \and
        \label{ins:sara2} Department of Aeronautics and Astronautics, MIT, 77 Massachusetts Avenue, Cambridge, MA 02139, USA
        \and
        \label{ins:nccr} NCCR/PlanetS, Centre for Space \& Habitability, University of Bern, Bern 3012, Switzerland\newpage
        \and
        \label{ins:tess1} Department of Physics and Kavli Institute for Astrophysics and Space Research, Massachusetts Institute of Technology, Cambridge, MA 02139, USA
        \and
        \label{ins:tess2} Department of Astrophysical Sciences, Peyton Hall, 4 Ivy Lane, Princeton, NJ 08544, USA
        \and
        \label{ins:daylan} Department of Physics and McDonnell Center for the Space Sciences, Washington University, St. Louis, MO 63130, USA
        \and
        \label{ins:tess4} NASA Goddard Space Flight Center, 8800 Greenbelt Road, Greenbelt, MD 20771, USA
        \and
        \label{ins:seti} SETI Institute, 189 Bernardo Ave., Suite 200, Mountain View, CA  94043, USA
        \and
        \label{ins:pappalardo} Pappalardo Fellow
        \and
        \label{ins:royal} Royal Astronomical Society, Burlington House, Piccadilly, London W1J 0BQ
        \and
        \label{ins:smith} Center for Astrophysics $\vert$ Harvard \& Smithsonian, 60 Garden Street, Cambridge, MA 02138, USA
        \and
        \label{ins:fred} Planetary Discoveries, Fredericksburg, VA 22405, USA 
             }

   \date{Received October 3, 2023; accepted January 12, 2024}

 
  \abstract
  {The Transiting Exoplanet Survey Satellite (\textit{TESS}) provides a continuous suite of new planet candidates that need confirmation and precise mass determination from ground-based observatories.
  This is the case for the G-type star TOI-1710, which is known to host a transiting sub-Saturn planet ($\mathrm{M_p}$\,$=$\,28.3\,$\pm$\,4.7\,$\mathrm{M}_\oplus$) in a long-period orbit ($P$\,=\,24.28\,d).
  Here we combine archival SOPHIE and new and archival HARPS-N radial velocity data with newly available \textit{TESS} data to refine the planetary parameters of the system and derive a new mass measurement for the transiting planet, taking into account the impact of the stellar activity on the mass measurement. We report for TOI-1710\,b a radius of $\mathrm{R_p}$\,$=$\,5.15\,$\pm$\,0.12\,$\mathrm{R}_\oplus$, a mass of $\mathrm{M_p}$\,$=$\,18.4\,$\pm$\,4.5\,$\mathrm{M}_\oplus$, and a mean bulk density of $\rho_{\rm p}$\,$=$\,0.73\,$\pm$\,0.18\,$\mathrm{g\,cm^{-3}}$, which are consistent at 1.2$\sigma$, 1.5$\sigma$, and 0.7$\sigma$, respectively, with previous measurements. Although there is not a significant difference in the final mass measurement, we needed to add a Gaussian process component to successfully fit the radial velocity dataset.  This work illustrates that adding more measurements does not necessarily imply a better mass determination in terms of precision, even though they contribute to increasing our full understanding of the system. Furthermore, TOI-1710\,b joins an intriguing class of planets with radii in the range 4--8\,$\mathrm{R}_\oplus$ that have no counterparts in the Solar System. A large gaseous envelope and a bright host star make TOI-1710\,b a very suitable candidate for follow-up atmospheric characterization.
   }


   \maketitle
%

\section{Introduction}

The Transiting Exoplanet Survey Satellite (\textit{TESS}; \citealp{TESS_Ricker}) is a NASA-sponsored space telescope launched on April 18, 2018.
The original \textit{TESS} mission was a two-year full-sky survey to search for transiting planets, but the mission obtained a first extension of observations covering from July 2020 to September 2022. \textit{TESS} is currently in its second extended mission, which will last at least until September 2025. One of the \textit{TESS} mission's main goals is to look for 50 small planets with $\mathrm{R_P}$<4\,$\mathrm{R}_\oplus$ suitable for atmospheric characterization.
However, during that search \textit{TESS} has detected planets with a wide range of radii.
In particular, TOI-216b (\citealp{TOI-216b}), TOI-421c (\citealp{TOI421c}), TOI-674b (\citealp{Murgas_TOI}), LTT\,9779b (\citealp{LTT9779b}), and TOI-257b (\citealp{TOI-257b}) are some examples of confirmed planets alerted by \textit{TESS} that have sizes between Uranus (4\,$\mathrm{R}_\oplus$) and Saturn (9.5\,$\mathrm{R}_\oplus$); they are called   sub-Saturns (as defined in \citealp{Petigura2017}).


Sub-Saturns are a very interesting group of exoplanets to study because they have no counterparts in our Solar System.
Their large sizes can be explained by a heavy metal core with a significant H/He envelope \citep{Lopez_2014}.
Since sub-Saturns are a lighter version of Jovian planets, they could potentially offer new benchmarks to study different envelope accretion scenarios for gas-giant planets.
Runaway accretion, suggested as a possible mechanism for the formation of Jupiter-like planets (\citealp{Pollack_1996, Hubickyj2005}), does not manage to explain these low-density planets; alternative processes have been proposed, such as accretion within a gas-depleted disk \citep{Lee_Accretion}.

As is shown in \citet{Petigura2017} and \citet{Nowak_2020}, systems holding sub-Saturn planets tend to present different characteristics and architectures, depending on whether they are single-planet or multi-planet systems.
Lone sub-Saturns are often more massive and orbit their host stars on more eccentric orbits than those in multi-planetary systems. These differences suggest that the presence of another planet (or planets) might play an important role in the formation of sub-Saturns.
Dynamical instabilities between planets could result in mergers or scattering to high-inclination orbits, establishing the observed architectures with high-mass planets in low-multiplicity systems (\citealp{Petigura2016, Petigura2017, HD89345b}).

TOI-1710\,b (\citealp{Konig_TOI1710}) is a transiting sub-Saturn planet on a 24-day orbit around a bright (V\,=\,9.5\,mag, J\,=\,8.3\,mag) G5-type  star. TOI-1710 (BD+76 227) is at a distance of 81\,pc and is located near the Camelopardalis constellation. \cite{Konig_TOI1710} used SOPHIE and HARPS-N radial velocities (RVs) to derive for TOI-1710\,b a radius of $\mathrm{R_p}$\,$=$\,5.34\,$\pm$\,0.11\,$\mathrm{R}_\oplus$ and a mass of $\mathrm{M_p}$\,$=$\,28.3\,$\pm$\,4.7\,$\mathrm{M}_\oplus$. They also explored the influence of the stellar activity of the host star on their dataset, and found no significant impact.

Here we combined the previously published RVs from \cite{Konig_TOI1710} with new HARPS-N observations to derive a refined mass measurement of TOI-1710\,b considering the impact of the stellar activity of its host star.
Furthermore, we included in our joint fit the new available \textit{TESS} photometric data to refine its orbital period, and other system parameters.

This paper is organized as follows. In Section\,\ref{Sec:Obs} we describe the observations used in this work. The stellar properties of TOI-1710 are reported in Section\,\ref{Sec: Stellar}. In Section\,\ref{Sec:Ana} we explain the methods used in the data analysis. The results and conclusions are discussed and presented in Section\,\ref{Sec: Dicussion}.


\section{Observations}
\label{Sec:Obs}

\subsection{\textit{TESS} photometry}

Listed as TIC\,445805961 in the \textit{TESS} Input Catalog (TIC; \citealp{Stassun2018}) and then later classified as the TESS object of interest TOI-1710, this star was observed by \textit{TESS} in 2 min cadence integrations in Sectors\,19, 20, 26, 40, 53, 59, and 60. 
It is not scheduled to   be observed again until Sector\,73, according to \textit{TESS}-point Web Tool.\footnote{\url{https://heasarc.gsfc.nasa.gov/wsgi-scripts/TESS/TESS-point_Web_Tool/TESS-point_Web_Tool/wtv_v2.0.py/}}

The \textit{TESS} raw data are reduced by the Science Processing Operations Center (SPOC; \citealp{SPOC}) at the NASA Ames Research Center and are publicly available at the Mikulski Archive for Space Telescopes (MAST\footnote{\url{https://mast.stsci.edu/portal/Mashup/Clients/Mast/Portal.html}}).
The light curves were extracted using simple aperture photometry (SAP; \citealp{SAP}) and corrected from systematics using the Presearch Data Conditioning (PDC) pipeline (\citealp{PDC_1, stumpe_2012, PDC_2}).
For this work we used the PDC-corrected SAP photometry, which is the \textit{TESS} product also used by \citet{Konig_TOI1710}.
The \textit{TESS} pixels considered to compute the  SAP and PDC-corrected SAP for Sectors\,19, 20, 26, 40, 53, 59, and 60 are shown in Figure\,\ref{Fig: TESS_TPF_S19_S20_S26}.
A single transit was detected in the Sector\,19 PDC-SAP flux time series, and the correct period can be identified in a combined search of the available sectors.

On February 19,  2020, the star was announced under the \textit{TESS} Object of Interest (TOI) number 1710 as a possible host for a transiting planet
in the TOI catalogue\footnote{\url{https://tev.mit.edu/data/}} (\citealp{Guerrero_2021}).
A signal with a period of $24.28$ days and a transit depth ($\Delta F$) of 3070.0\,ppm, corresponding to a planet radius of about 5.4\,$\mathrm{R}_\oplus$, was detected in the Quick Look Pipeline (QLP; \citealp{QLP_1, QLP_2} ).
All SPOC data validation (\citealp{Twicken2018, Li_2019}) diagnostic test results for this signal are fully consistent with a transiting planet associated with the host star (within 0.7\,$\pm$\,2.6$"$) and excluded   all other TIC objects.

\subsection{Ground-based photometry}
\label{subsec: ground phot}

We observed a full transit of TOI-1710.01 on the night of February 29, 2020, with the George Mason University Observatories 0.8 m telescope.
We used the \textit{TESS} Transit Finder to schedule our transit observation (\citealp{tapir}).
We utilized \texttt{AstroImageJ} (\citealp{AstroImageJ}) for data reduction, plate-solving, aperture photometry, light curve extraction, and detrending.
Due to differences in quality and scattering between space- and ground-based measurements, we analysed the two datasets individually, focusing our photometric analyses on the \textit{TESS} light curves, and we did not include these observations in the joint fit modelling in Sect.\,\ref{subsec: JF model}.
However, ground-based observations are key to the independent confirmation and validation of \textit{TESS} candidates.

\subsection{High-resolution spectroscopy with SOPHIE}
\label{subsec: SOPHIE}

\cite{Konig_TOI1710} performed a ground-based  follow-up campaign with the SOPHIE spectrograph. They obtained a total of 30 RV measurements acquired from September 14, 2020, to May 7, 2021. The details of these observations are explained in Sect.\,2.2.2 of \cite{Konig_TOI1710}.

\subsection{High-resolution spectroscopy with HARPS-N}
\label{subsec: HARPS-N}

TOI-1710 was observed with the High Accuracy Radial velocity Planet Searcher for the Northern hemisphere (HARPS-N; \citealp{HARPS-N}) mounted on the 3.6m \textit{Telescopio Nazionale Galileo} (TNG) at the Roque de los Muchachos Observatory, La Palma. The star was monitored first from October 27,  2020, to April 5, 2021, and then from September 12, 2021, to April 4,  2022. We obtained a total of 54 high-resolution ($R \sim 115\,000$) spectra. The observations were carried out as part of observing programs CAT20B\_41 and CAT21A\_24 (PI: Pall\'e). The exposure times varied from 680 to 2400 seconds, depending on weather conditions and scheduling constraints, leading to a S/N per pixel of 33--106 at 5500\,\AA. The spectra were extracted using the offline version of the HARPS-N DRS pipeline \citep{2014SPIE.9147E..8CC}, version 3.7. Doppler measurements (absolute RVs) and spectral activity indicators (cross-correlation function full width at half maximum, FWHM; cross-correlation function contrast, CTR; cross-correlation function bisector, BIS; and $R_{\rm HK}$) were measured using an online version of the DRS, the YABI tool,\footnote{Available at \url{http://ia2-harps.oats.inaf.it:8000}.} by cross-correlating the extracted spectra with a G2 mask \citep{1996A&AS..119..373B}. We also used the {\tt serval} code \citep{SERVAL}  to measure relative RVs by template-matching, chromatic index (CRX), differential line width (dLW), and H$\alpha$, sodium Na\,D1 and Na\,D2 indexes. The uncertainties of the RVs measured with {\tt serval} are in the range 0.8--2.6\,$\mathrm{m\,s^{-1}}$, with a mean value of 1.4\,$\mathrm{m\,s^{-1}}$. The uncertainties of absolute RVs measured with DRS are in   the range 0.8--3.2\,$\mathrm{m\,s^{-1}}$, with a mean value of 1.5\,$\mathrm{m\,s^{-1}}$. Table\,\ref{table: table_RV_1} gives the time stamps of the spectra in BJD$_{\mathrm{TDB}}$ {\tt serval} relative RVs along with their $1\sigma$ error bars, and spectral activity indicators measured with YABI and {\tt serval}.

\cite{Konig_TOI1710} also performed a ground-based follow-up campaign with HARPS-N, and  obtained 31 measurement acquired from October 3, 2020, to April 19, 2021. \cite{Konig_TOI1710} measurements from SOPHIE and HARPS-N are contemporaneous with our first observing campaign. Moreover, there are no significant differences between the two HARPS-N datasets: consistent exposure times, similar S/N per pixel, and measured RV uncertainties, and the same instrumental mode. For consistency, we reduced their HARPS-N spectra together with our HARPS-N observations, with the YABI tool and the HARPS-N DRS pipeline. The RV values (from the  \citealt{Konig_TOI1710} spectra) extracted using the DRS pipeline are exactly the same  as those presented in \citet[Table\,C.2]{Konig_TOI1710}. We ran the \texttt{serval} pipeline over both datasets, considering them  a single dataset of 85 HARPS-N RVs. The RVs extracted by \texttt{serval} are used in the analyses of this work.

\subsection{High-resolution imaging with Gemini/`Alopeke}
\label{subsec: Gemini}

\begin{figure}
    \centering
    \includegraphics[width=\hsize]{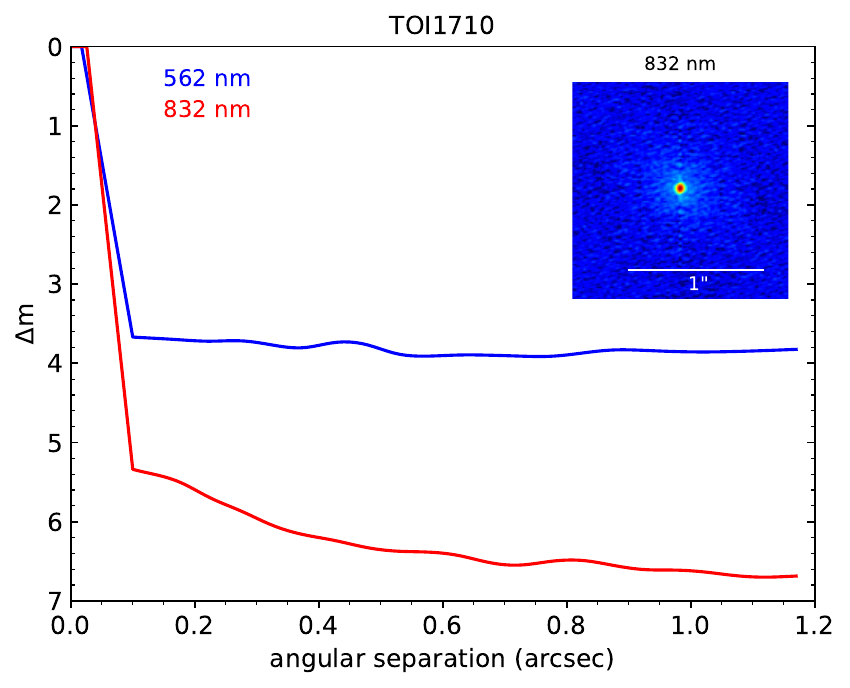}
    \caption{ Gemini/`Alopeke high-resolution image of TOI-1710 take on   February 9, 2021. TOI-1710 is a single star to contrast limits of 5.5 to 7\,mag within 1.6\,AU out to 97\,AU (d\,=\,81\,pc). The inset shows a  1.2"$\times$1.2"   reconstructed image of TOI-1710 at 832\,nm band, centred on the star.}
    \label{Fig: HR Gemini}
\end{figure}

If an exoplanet host star has a spatially close companion, that companion (bound or line of sight) can create a false-positive transit signal if it is, for example, an eclipsing binary.
Additionally, `third-light' flux from the close companion star can lead to an underestimated planetary radius if not accounted for in the transit model \citep{Ciardi_2015} and cause non-detections of small planets residing within the same exoplanetary system \citep{Lester_2021_new}.
The discovery of close-in bound companion stars, which exist in nearly one-half of FGK-type stars \citep{Matson_2018}, provides crucial information  towards our understanding of exoplanetary formation, dynamics, and evolution \citep{Howell_2021}.
Thus, to search for close-in bound companions, unresolved in \textit{TESS} or other ground-based follow-up observations, we obtained high-resolution imaging speckle observations of TOI-1710.

TOI-1710 was observed on  February 9, 2021,  using the ‘Alopeke speckle instrument on the Gemini North 8 m telescope.\footnote {\url{https://www.gemini.edu/sciops/instruments/alopeke-zorro/}}
‘Alopeke provides simultaneous speckle imaging in two bands (562\,nm and 832\,nm) with output data products including a reconstructed image with robust contrast limits on companion detections  (e.g. \citealp{Howell_2016}).
Seven sets of 1000$\times$0.06\,sec exposures were collected and subjected to Fourier analysis in our standard reduction pipeline (see \citealp{Howell_2011}).
Figure\,\ref{Fig: HR Gemini} shows our final contrast curves and the 832\,nm reconstructed speckle image.
We find that TOI-1710 is a single star with no companion brighter than 5.5--7 magnitudes below that of the target star from the diffraction limit (20\,mas) out to 1.2”.
At the distance of TOI-1710 (d\,=\,81\,pc) these angular limits correspond to spatial limits of 1.6 to 97\,AU.

\section{Stellar system analysis}
\label{Sec: Stellar}

\begin{table}
\caption{\label{table - stellar parameters} Stellar parameters of TOI-1710.}
\centering
\resizebox{\columnwidth}{!}{%
\begin{tabular}{lcr}
\hline
\hline
\noalign{\smallskip} 
Parameter & Value & Reference \\
\noalign{\smallskip} 
\hline
\noalign{\smallskip} 
\multicolumn{3}{c}{Name and identifiers} \\
\noalign{\smallskip} 

TIC & 445805961 & \textit{TESS}\\
TOI & 1710 & TOI\\
BD & +76 227 & BD \\
TYC & 4525--1009--1 & TYC \\
2MASS & J06170789+7612387 & 2MASS\\
Gaia DR2 & 1116613161053977472 & \textit{Gaia}\\
\noalign{\smallskip} 
\hline
\noalign{\smallskip} 
\multicolumn{3}{c}{Coordinates and spectral type} \\
\noalign{\smallskip} 

$\alpha$\,(J2000) & 06$^\mathrm{h}\,$17$^\mathrm{m}$\,07$^\mathrm{s}\!$.86 & \textit{Gaia} \\
$\delta$\,(J2000) & $+$76º\,12$'$\,38$"\!$.81 & \textit{Gaia} \\
Spectral type& G5\,V & W03 \\

\noalign{\smallskip} 
\multicolumn{3}{c}{Parallax and kinematics} \\
\noalign{\smallskip} 

$\pi$ ~[mas] & 12.2823 $\pm$ 0.0266 & \textit{Gaia} \\
$d$ ~[pc] & 81.42  $\pm$  0.18 & \textit{Gaia} \\
$\mu_{\alpha} \cos{\delta}$ ~[mas\,year$^{-1}$] & 59.837 $\pm$ 0.037 & \textit{Gaia} \\
$\mu_\delta$ ~[mas\,year$^{-1}$] & 55.610 $\pm$ 0.044 & \textit{Gaia} \\
$V_{\rm r}$ ~[$\mathrm{km\,s^{-1}}$] & $-$39.40 $\pm$ 0.35 & \textit{Gaia} \\
              & $-$38.8134 $\pm$ 0.0007 & Sec.\,\ref{subsec: HARPS-N}  \\

\noalign{\smallskip} 
\multicolumn{3}{c}{Magnitudes} \\
\noalign{\smallskip} 

B [mag] & 10.20 $\pm$ 0.03 & TYC \\
V [mag] & 9.54 $\pm$ 0.02  & TYC \\
G [mag] & 9.3598 $\pm$ 0.0002 & \textit{Gaia} \\ 
T [mag] & 8.9134 $\pm$ 0.006 & \textit{TESS} \\
J [mag] & 8.319 $\pm$ 0.019 & 2MASS \\
H [mag] & 8.003 $\pm$ 0.034 & 2MASS \\
K [mag] & 7.959 $\pm$ 0.026 & 2MASS \\

\hline
\noalign{\smallskip} 
\multicolumn{3}{c}{Stellar parameters} \\

Mass ~$M_{\star}$ [$M_{\odot}$] & 0.99 $\pm$ 0.07 & Sec.\,\ref{subsec: Params} \\
 & 0.984$^{+0.050}_{-0.059}$ & K22 \\

Radius ~$R_{\star}$ [$R_{\odot}$] & 0.95 $\pm$ 0.02 & Sec.\,\ref{subsec: Params} \\
  & 0.968$^{+0.016}_{−0.014}$ & K22 \\

Luminosity ~$L_{\star}$ [$L_{\odot}$]  & 0.895 $\pm$ 0.003 & \textit{Gaia} \\

Effective temperature ~$T_{\rm eff}$ [K] & 5730 $\pm$ 30 & Sec.\,\ref{subsec: Params} \\
 & 5665 $\pm$ 55 & K22 \\

Surface gravity ~$\log$(g) & 4.54 $\pm$ 0.09 & Sec.\,\ref{subsec: Params} \\
  & 4.46 $\pm$ 0.10  & K22\\

Metallicity ~$\rm [Fe/H]$ & 0.12 $\pm$ 0.06 & Sec.\,\ref{subsec: Params} \\
 &  0.10 $\pm$ 0.07  & K22\\
Age ~[Gyr] & 2.8 $\pm$ 0.6 & Sec.\,\ref{subsec: Rotation} \\
 & 4.2$^{+4.1}_{-2.7}$  & K22\\
log(R$^{'}_{HK}$) & $-$4.786 $\pm$ 0.012 & Sec.\,\ref{subsec: Rotation} \\
 &  $-$4.78 $\pm$ 0.03  & K22\\
\noalign{\smallskip}
\hline

\end{tabular}
}

\tablebib{\textit{TESS}: \cite{Stassun2018, Stassun2019}, TOI: \citet{Guerrero_2021}, BD: \citet{BD_catalog}, TYC: \cite{TYC_cat}; \textit{Gaia}: \cite{GaiaColab, GAIA_DR3}; 2MASS: \cite{2MASS_Cat}; W03: \cite{Wright-2003}; K22: \cite{Konig_TOI1710}. }


\end{table}

\subsection{Stellar companions}
\label{subsec: companions}

Given the fainter TOI-1710 companion  labelled  $\#$2 in Figure\,\ref{Fig: TESS_TPF_S19_S20_S26} and the large \textit{TESS} pixel size of 21", it is crucial to verify that no visually close-by targets are present.
In Section\,\ref{subsec: Gemini} we ruled out this scenario within 1.2" around TOI-1710 (see Fig.\,\ref{Fig: HR Gemini}). 
To go wider, we extended the exploration of close companions until 60" around the star position using the \textit{Gaia} DR2 (\citealp{GaiaColab}) database.
The main astrometric parameters and properties of the five found objects, relative to TOI-1710, are included in Table\,\ref{table - gaia companions}.

\textit{Gaia} DR2\,1116613161052594560 at 7.2" with G$_{RP}$\,=\,17.4\,mag is the only target within one \textit{TESS} pixel, but it is much fainter than TOI-1710 (G$_{RP}$\,=\,8.9\,mag).
Stars with differences in \textit{Gaia} G$_{RP}$-band larger than 8\,mag are not detected in the \textit{TESS} images (Figure\,\ref{Fig: TESS_TPF_S19_S20_S26}).
Only \textit{Gaia} DR2\,1116612783096856960, which has $\Delta$\,G$_{RP}$\,=\,4.2\,mag, is detected. We identified this star as TIC\,445805957, which is the southern star  labelled  $\#$2 in Figure\,\ref{Fig: TESS_TPF_S19_S20_S26}.
Since star $\#$2 has a parallax and proper motion similar to that of  TOI-1710, we can assume that  they are probably gravitationally bounded.
This is in agreement with \citet{Mason_binary} and \citet{Tian2020}, who reported this pair of stars as a wide binary system.

Because the \textit{Gaia} G$_{RP}$-band (630--1050\,nm) and the \textit{TESS} band (600--1000\,nm) are very much alike, we could 
estimate the dilution factor for the \textit{TESS} photometry (D$_{\textit{TESS}}$) using its definition from Eq.\,6 in \citet{juliet}.
We calculated a D$_{\textit{TESS}}$\,=\,0.988 considering all the nearby stars, corresponding to 1.2\% flux contamination.
However, the PDC-corrected SAP light curve, which is the data used in the photometry and joint fit analyses (Sections\,\ref{subsec: lc model} and \ref{subsec: JF model}), already takes into account the possible flux contamination by the nearby companion, which is by far the brightest close star.
Excluding the flux from TIC\,445805957 from the calculations, D$_{\textit{TESS}}$ is 0.999 (i.e. consistent with little to no flux contamination).

\begin{table}
\caption{\label{table - gaia companions} Relative properties to TOI-1710 of the nearby stars. TOI-1710 wide binary companion \textit{Gaia} DR2 identifier is marked in bold.
}
\centering
\resizebox{\columnwidth}{!}{%
\begin{tabular}{ccccc}
\hline
\hline
Gaia & Separation & $\Delta$\,Parallax & $\Delta$\,Proper motion & $\Delta$\,G$_{RP}$ \\
DR2 & [\,"\,] & [\,mas\,] & [\,mas/year\,] & [\,mag\,] \\

\hline


1116613161052594560 & 7.2\,$\pm$\,0.1 & 12.0\,$\pm$\,0.2 & 72.66\,$\pm$\,0.14 & 8.545\,$\pm$\,0.025 \\
1116613225476795776 & 30.8\,$\pm$\,0.1 & 12.30\,$\pm$\,0.25 & 79.35\,$\pm$\,0.14 & 9.153\,$\pm$\,0.020 \\
1116613195412077696 & 37.1\,$\pm$\,0.1 & 10.9\,$\pm$\,0.7 & 72.2\,$\pm$\,0.5 &  10.292\,$\pm$\,0.050 \\
1116613225476796288 & 42.1\,$\pm$\,0.1 & 12.1\,$\pm$\,0.3 & 80.02\,$\pm$\,0.17 & 9.592\,$\pm$\,0.020 \\
{\bf 1116612783096856960} & 43.3\,$\pm$\,0.1 & 0.0\,$\pm$\,0.06 & 0.087\,$\pm$\,0.003 & 4.2307\,$\pm$\,0.0018 \\

\hline

\end{tabular}
}


\end{table}

\subsection{Stellar rotation}
\label{subsec: Rotation}


Based on the  \citet{1984ApJ...279..763N} and \citet{2008ApJ...687.1264M} activity-rotation relations and using (B-V) of 0.658 and the $\log\,R^{'}_{HK}$ measured with YABI, we estimated a rotation period of TOI-1710 for 20.9\,$\pm$\,4.2\,days and 20.6\,$\pm$\,3.2\,days, respectively. Using the activity-age relation of \citet{2008ApJ...687.1264M}, we also found the age of TOI-1710 to be in the range of 2.2--3.9\,Gyr.

We searched for photometric time series from automated ground-based surveys to detect long-term photometric modulations associated with the stellar rotation.
The Northern Sky Variability Survey (NSVS; \citealp{NSVS}) is the only public survey in which the search was successful.
The search parameters were the TOI-1710 coordinates from Table\,\ref{table - stellar parameters} and a radius of 1' around it.

The query returned two objects labelled with different identifiers: ($i$) $\#$594837 (RA\,=\,06:17:07.86, Dec\,=\,+76:12:38.23, and magnitude\,=\,9.657\,$\pm$\,0.011\,mag with a scatter magnitude of 0.024\,mag) and ($ii$) $\#$628193 (RA\,=\,06:17:07.88, Dec\,=\,+76:12:38.41, and magnitude\,=\,9.671\,$\pm$\,0.011\,mag with a scatter magnitude of 0.021\,mag). Objects
$\#$594837 and $\#$628193 are only separated from each other by 0.20" and they are 0.58" and 0.41" apart from TOI-1710, respectively.
Their magnitudes are coincident within the scatter magnitudes and they are only 0.1\,mag apart from TOI-1710 V-band magnitude. Since we already confirmed in Section\,\ref{subsec: companions} that there is no other star of similar magnitude within 60" around TOI-1710, we can safely assume that the  NSVS observations of $\#$594837 and $\#$628193 correspond to the same object, TOI-1710.

\begin{figure}
    \centering
    \includegraphics[width=\hsize]{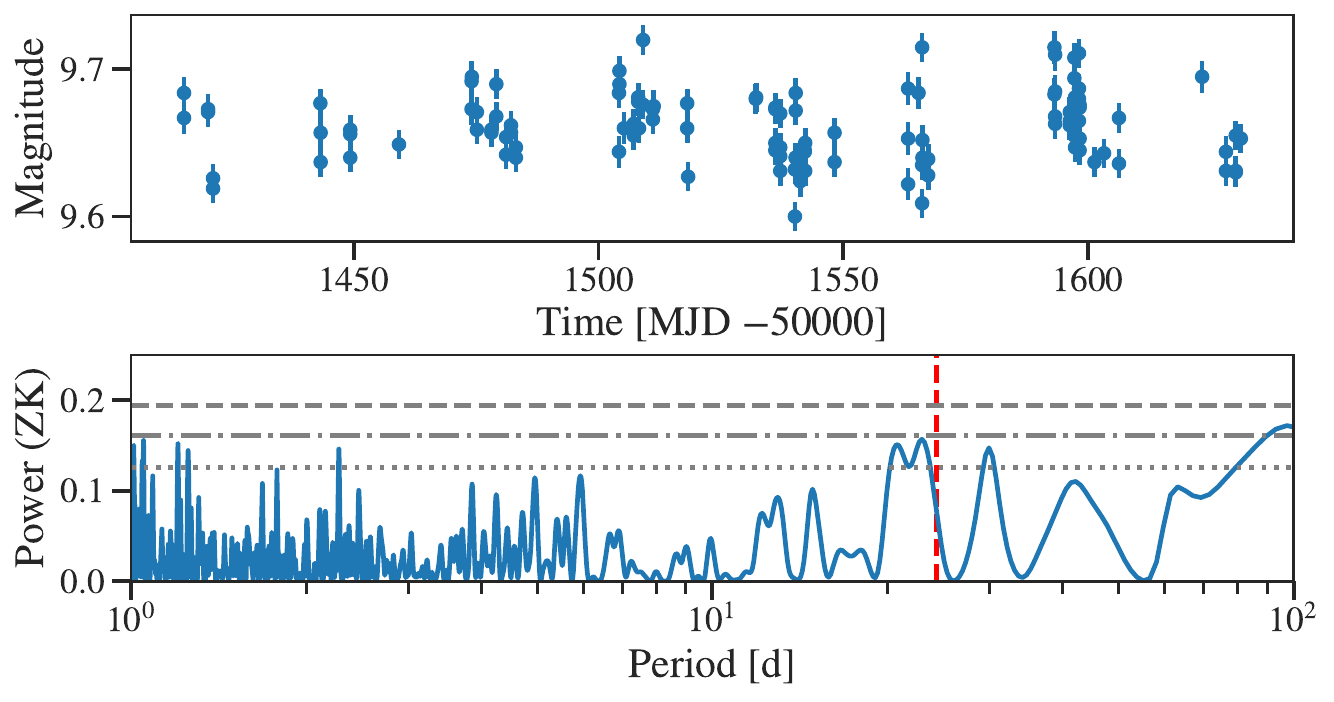}
    \caption{ Dataset (\textit{top}) and  generalized Lomb-Scargle periodogram (\textit{bottom}) of Northern Sky Variability Survey photometric data for TOI-1710. The vertical red dashed line marks the 24.28\,d period. The 10\%, 1\%, and 0.1\% FAP levels are indicated by horizontal grey dotted, dash-dotted, and dashed lines, respectively.}
    \label{Fig:NSVS}
\end{figure}

We searched for photometric periodicities in the total of 117 NSVS measurements using the generalized Lomb-Scargle (GLS) periodogram\footnote{\url{https://github.com/mzechmeister/GLS}} \citep{GLS_paper}. In addition, we computed the theoretical false alarm probability (FAP), as described in \citet{GLS_paper}. The photometry and its GLS periodogram are shown in Figure\,\ref{Fig:NSVS}.
A baseline of 216 days should be enough to detect any significant photometric modulation in the expected range for the stellar rotational period.
The only relevant signal is a broad double peak in the range 20--24\,d with a FAP close to the 1\%, and consistent with the $P_{\rm rot}$ reported by \cite{Konig_TOI1710} of $\sim$22\,d.

\subsection{Stellar parameters}
\label{subsec: Params}

The analysis of the stellar spectrum was carried out by using the \texttt{BACCHUS} code \citep{2016ascl.soft05004M} relying on the MARCS model atmospheres \citep{2008A&A...486..951G} and using the co-added HARPS-N spectra. In brief, an effective temperature of 5730\,$\pm$\,10\,K was derived  by requiring no trend of the \ion{Fe}{I} lines abundances against their respective excitation potential. The surface gravity was determined by requiring the  ionization balance of the  \ion{Fe}{I}  and \ion{Fe}{II} lines. A microturbulence velocity value of 0.89\,$\pm$\,0.05\,$\mathrm{km\,s^{-1}}$ was also derived by requiring no trend of Fe line abundances against their equivalent widths. The output metallicity ($+0.12$\,$\pm$\,0.06) is represented by the average abundance of the \ion{Fe}{I} lines. 
An alternative analysis of the HARPS-N spectrum was done with the SPC pipeline \citep{2012Natur.486..375B}. The results obtained with this pipeline are  $T_{\rm eff}$\,=\,5720\,$\pm$\,28, $\log g$\,=\,4.49\,$\pm$\,0.04, and $\rm [M/H]$\,=\,$+$0.06\,$\pm$\,0.04. These results are consistent with the parameters derived with the \texttt{BACCHUS} pipeline.
Nevertheless, the internal error in temperature that was  derived for example by the \texttt{BACCHUS} code (10\,K) or the SPC code (28\,K) are unrealistically small as they do not include systematic errors. One of the main known sources of uncertainties are NLTE or 3D effects that are not considered in either of the 1D/LTE analyses of the HARPS-N spectrum. 
In order to try to better estimate its global error, the effective temperature was also derived with photometry. While the spectroscopic method we  employed relies on individual line strengths of the observed stellar spectrum, the photometric method relies on the relative flux of the star and its observed colours. In that sense, one can consider that the spectroscopic and the photometric method provide fairly independent results, and thus better reflect the absolute uncertainty on temperature. Nevertheless, the photometric method also has its own uncertainties. To minimize the error, we chose the V-K colour index.  This index has been demonstrated to be the most robust and most reliable temperature indicator because it offers a wider baseline of the spectral energy distribution and because it is less dependent on other parameters, such as metallicity and surface gravity \citep{1998A&A...333..231B,2010A&A...512A..54C}. Another known source of uncertainty of the photometric method resides in the absolute calibration  that sets the zero-point of the $T_{\rm eff}$ scale; this calibration varies from one study to another. 
By using the V-K colour-temperature relation of \citet{2009A&A...497..497G} and that of \citet{2010A&A...512A..54C} and assuming no reddening for such a close-by star, we  obtained a temperature of 5669\,$\pm$\,32\,K and 5719\,$\pm$\,25\,K, respectively. While all photometric and spectroscopic temperatures are in good agreement, we chose to compute the error in temperature by taking   half of the difference between the most extreme spectroscopic and photometric temperatures, which are  the \texttt{BACCHUS} and the \citet{2009A&A...497..497G} temperatures, leading to a total error of 30\,K.  

In a second step, we used the Bayesian tool PARAM \citep{2014MNRAS.445.2758R,2017MNRAS.467.1433R} to derive the stellar mass, radius, and age utilizing the spectroscopic parameters and the updated \textit{Gaia} luminosity along with our spectroscopic temperature. However, these  Bayesian tools underestimate the error budget as they do not take into account the systematic errors between one set of isochrones to another, due to the various underlying assumptions in the respective stellar evolutionary codes. In order to take into account those systematic errors, we combined the results of the two sets of isochrones provided by PARAM (i.e. MESA and Parsec) and added the difference between the two sets of results to the error budget provided by PARAM. We obtained a stellar radius and mass of   $\rm 0.95 \pm 0.02\,R_\odot$ and $\rm 0.99 \pm 0.07\,M_\odot$, respectively. We note, however,  that although using two sets of isochrones may mitigate underlying systematic errors, our formal error budget for radius and luminosity may still be underestimated, as demonstrated by \citet{Tayar2022}. For solar-type stars such as TOI-1710, absolute errors may instead be up to 4\%, 2\%, 5\%, and 20\%   respectively for radius, luminosity, mass, and age.

\begin{figure}
    \centering
    \includegraphics[width=\hsize]{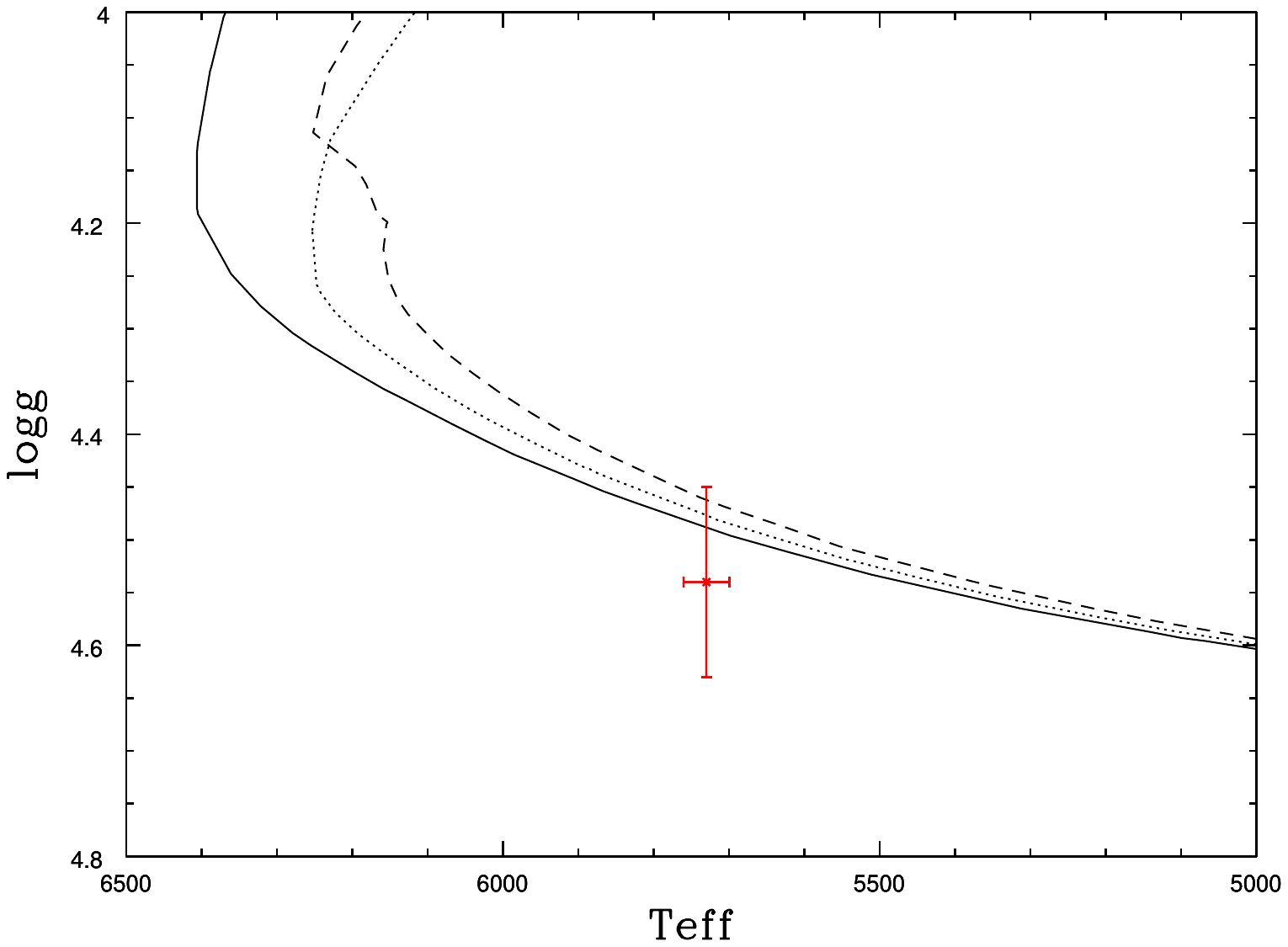}
    \caption{Location of the host star in the spectroscopic H-R diagram. The red point and error bars indicate the spectroscopic parameters of the host star, while the solid, short-dashed, and long-dashed lines show isochrones for 2.5, 3.0, and 4.0\,Gyr, respectively. }
    \label{Fig:HR}
\end{figure}

The derived age from the isochrones (3.2\,$\pm$\,3.1\,Gyr) do not provide constraining information as age is very degenerate for stars on the main sequence (see Fig.\,\ref{Fig:HR}) . Therefore, we rely on the study of \citet{2008ApJ...687.1264M} for the choice of the age (2.8\,$\pm$\,0.6\,Gyr) derived from the empirical R$^{'}_{HK}$-age relation.

All stellar parameters adopted in our joint modelling of the system (presented in Section~\ref{Sec:Ana}) are summarized in Table\,\ref{table - stellar parameters}. In Table\,\ref{table - stellar parameters} are also displayed the values provided by   \cite{Konig_TOI1710}. The values are in good agreement, although we derive a slightly higher temperature. The method employed to derive the temperature in \cite{Konig_TOI1710} is very similar to ours, and the small discrepancy is likely due to the differences in the model atmosphere, radiative transfer codes, and the selection of Fe lines, as demonstrated in \citet{Jofre2014}. While such a difference in temperature consistently leads to a slightly higher mass and smaller radius between the two studies, $M_\star$, $R_\star$, log(g), metallicity, age, and log(R$^{'}_{HK}$) are still in agreement within the error bars.

\section{Analysis}
\label{Sec:Ana}

\subsection{Frequency analysis and RV correlations}

\begin{figure}
  \centering
  \includegraphics[width=0.49\textwidth]{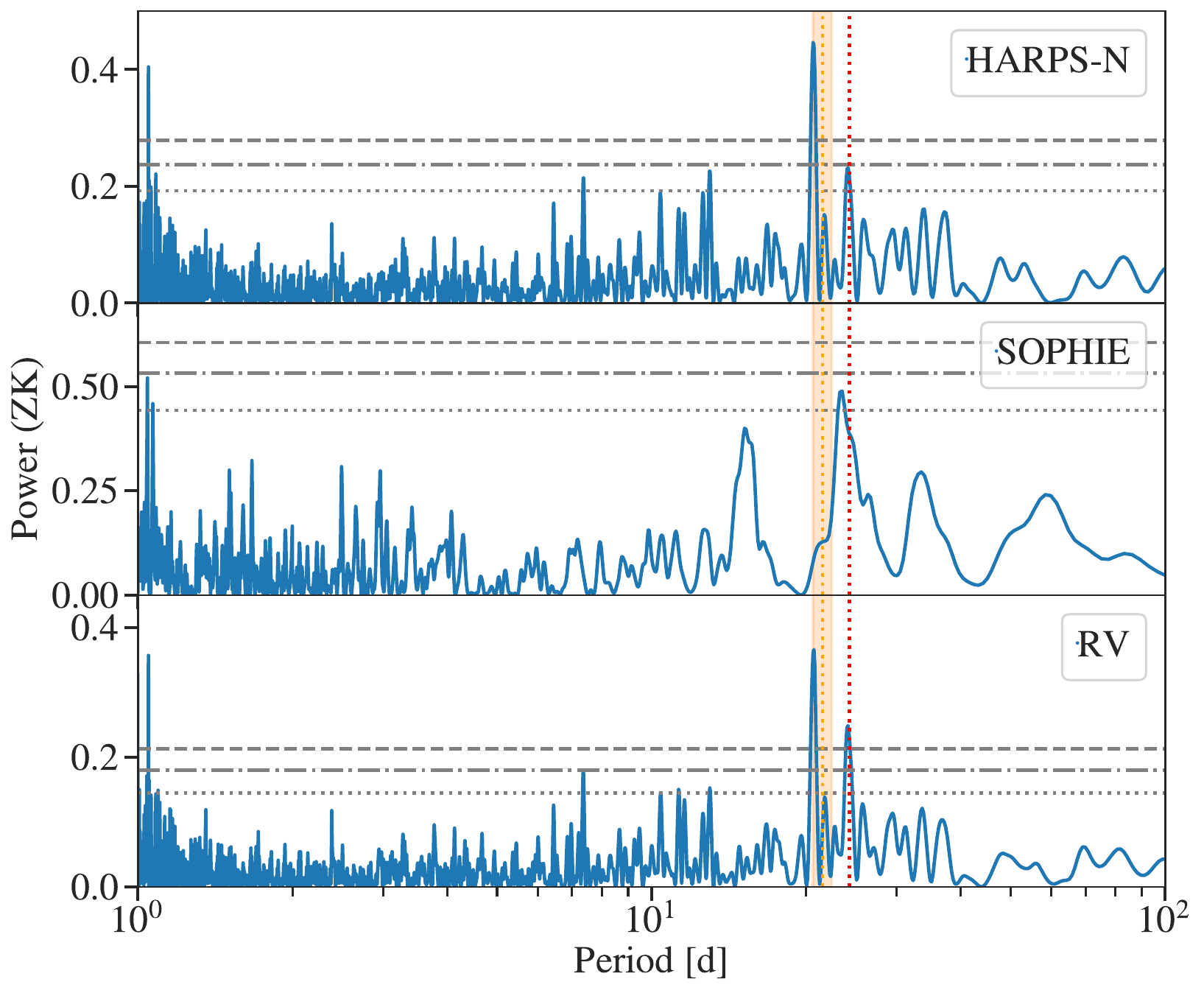}
  \caption{Generalized  Lomb-Scargle periodograms of HARPS-N, SOPHIE, and combined RV measurements (from top to bottom). Vertical red dotted line indicates the period of TOI-1710\,b at 24.28\,d. Shaded orange region indicates the 3$\sigma$ region of the $P_{\rm rot}$ from the joint fit (see Sect.\,\ref{subsec: JF model}).
  The graph shows 10\%, 1\% and 0.1\% FAP levels in grey dotted, dash-dotted, and dashed, respectively. \label{fig: GLS RVs}}
\end{figure}

Our RV dataset involves a total of 115 measurements from HARPS-N (85) and SOPHIE (30) spectrographs with a time baseline of 217 days.
We searched for TOI-1710\,b and other signals in the HARPS-N, SOPHIE, and combined RV datasets computing their GLS periodograms, which are shown in Figure\,\ref{fig: GLS RVs}. The signal of the planet is well detected in the RV datasets with FAP $\lesssim$\,1\,\%. However, the biggest peak in the HARPS-N and the combined dataset periodograms is at $\sim$20\,days, which is consistent with the rotation period of the star derived in Sect.\,\ref{subsec: Rotation} and by \cite{Konig_TOI1710}. Thus, the RVs analysed in this work have two clear signals that  have to be modelled: the transiting planet TOI-1710\,b at 24.28\,d, and the stellar rotation at 20--22\,d. Due to the proximity of the two signals, they should be treated with caution when analysing the RVs.

We computed the GLS periodogram of the HARPS-N activity parameters measured with YABI and \texttt{serval} (see Figure\,\ref{Fig:ACT}). In general, all the activity periodograms present peaks in the region where the stellar rotation is expected (20--24\,d). These peaks are significant (FAP\,<\,1\,\%) for  CTR, FWHM, dLW, Na\,D1, and $R_{\rm HK}$. All the activity indicators except CRX and Na\,D2 also present significant peaks at $\sim P_{\rm rot}/2$. \citet{Konig_TOI1710} also presented the GLS periodograms for FWHM, H$\alpha$, and log(R`$_{HK}$) with significant peaks close to the stellar rotation period.
Moreover, the HARPS-N CTR, FWHM, dLW, and H$\alpha$ index have strong correlations with the HARPS-N RVs (see Fig.\,\ref{Fig: correlation}). Similar correlations are obtained when we split the HARPS-N dataset in two (before and after September 2021), indicating that the stellar correlation did not come (only) from the second epoch and that it is present over all the observations. The stellar rotation detected in the RV GLS periodogram and the correlation of the  RVs with the FWHM, for example, are not negligible.

\subsection{Photometric modelling}
\label{subsec: lc model}

In a first step, previous to the RV and photometry simultaneous analysis, we studied the photometric data alone.
Although TOI-1710 was observed in Sectors\,19, 20, 26, 40, 53, 59, and 60, the TOI-1710\,b expected transit in Sector\,60 fell in the mid-sector observing gap. Thus, we did not consider Sector\,60 in our analysis because it has no 1710\,b transits, and to save computational time. We analysed the \textit{TESS} data from Sectors\,19, 20, 26, 40, 53, and 59 with the \texttt{Python} library \texttt{juliet} (\citealp{juliet}). \texttt{Juliet} performs the fitting procedure using other public packages for modelling of transit light curves (\texttt{batman}, \citealp{batman}) and GPs (\texttt{celerite}, \citealp{celerite}). Instead of using the Markov chain Monte Carlo (MCMC) technique, \texttt{juliet} uses a nested sampling algorithm (\texttt{dynesty}, \citealp{dynesty}; \texttt{MultiNest}, \citealp{MultiNest, PyMultiNest}) to explore the whole the parameter space.
We considered the uninformative sample ($r_1$,$r_2$) parametrization  introduced in \citet{Espinoza2018} to explore the impact parameter of the orbit ($b$) and the planet-to-star radius ratio ($p$\,$=$\,${R}_{\mathrm{p}}/{R}_{\star}$) values. In the fitting procedure we adopted a quadratic limb-darkening law with the ($q_1$,$q_2$) parametrization introduced by \citet{Kipping2013}. According to the flux contamination exploration in Sect.\,\ref{subsec: companions}, we can safely fix the dilution factor to 1. We added  a \texttt{celerite} GP exponential kernel to the transiting planet
model to account for the extra correlated noise in the photometric data.

The fitted parameters with their prior and posterior values, and the derived parameters are shown in Table\,\ref{table - juliet priors and posteriors}. The \textit{TESS} data along with the transiting and GP models are shown in Fig.\,\ref{Fig: juliet TESS SECTORS}. The TOI-1710\,b phase-folded  photometry from the  \texttt{juliet} anaysis is shown in Fig.\,\ref{Fig: juliet PHASE FOLDED}. The results from the photometric data analysis are consistent with the planetary characteristics from \cite{Konig_TOI1710}.

Furthermore, we analysed the ground-based TOI-1710\,b transit modelling the airmass-detrended light curve with \texttt{ExoFASTv2} (\citealp{ExoFast}) to validate the detection.
The independent analysis confirmed the results derived from the \textit{TESS} light curves, and we recovered a $8\sigma$ significance and on-time and on-target transit with consistent depth, duration, and timing with the \textit{TESS} ephemerides.
Ground-based photometry along with its transit fit is shown in Figure\,\ref{Fig: GB Photometry}.

\subsection{Joint fit modelling}
\label{subsec: JF model}

\begin{figure}
  \centering
  \includegraphics[width=0.49\textwidth]{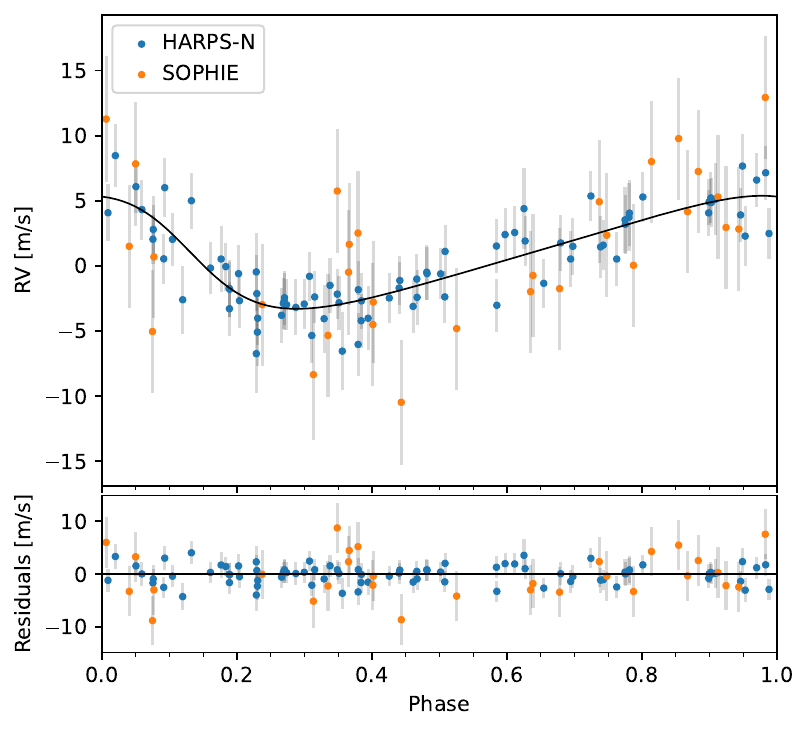}
  \includegraphics[width=0.49\textwidth]{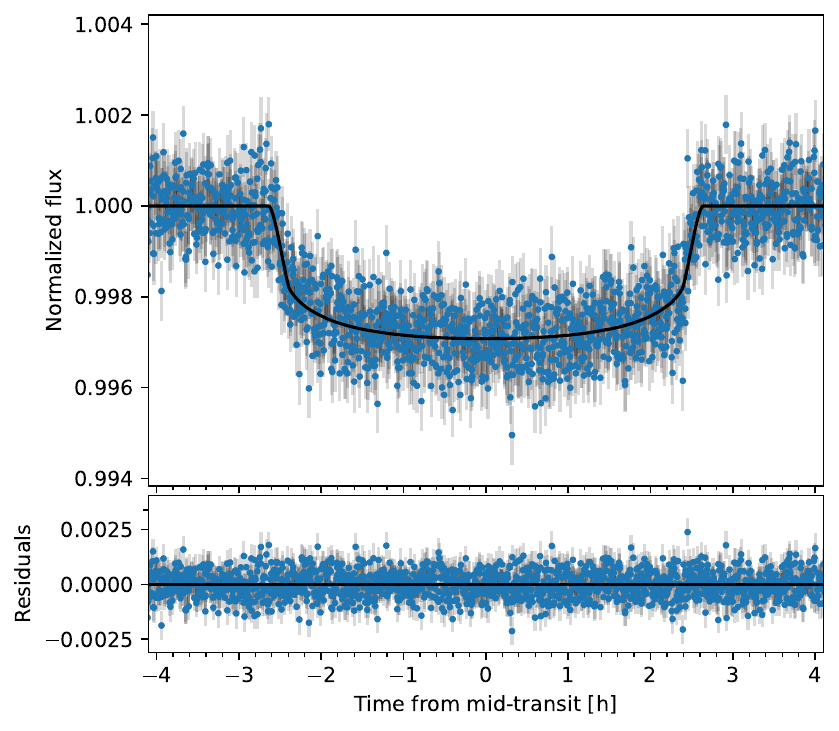}
  \caption{
  Phase-folded RV (top) and TESS (bottom) data for TOI-1710\,b.
 \textit{Upper panel}: HARPS-N (blue) and SOPHIE (orange) RV data with the overplotted fit from the 1pGP model. \textit{Lower panel}: Combination of \textit{TESS} transits from Sectors 19, 20, 26, 26, 40, 53, and 59. The black line is the inferred 1pGP model. \label{fig:joint_fit_plot}}
\end{figure}

\begin{table}
\centering
\caption[width=\columnwidth]{ Statistical criteria for the three tested models. Listed are the model names, BIC and AIC values, and  number of free parameters. The best criteria model is marked in bold. \label{tab:bicaic}}
\begin{tabular}{lccc}
\hline
Model &     BIC & AIC  &  $N_\mathrm{free}$  \\
\hline

1p & $-$61120.8 & $-$61370.8 &  38\\
1pGP & $-$61116.9 & $-$61399.8  &  43  \\
\textbf{1pGP+FWHM} & \textbf{$-$61340.0} & \textbf{$-$61643.4} &  \textbf{46}  \\
\hline
\end{tabular}
\end{table}

\begin{table*}
  \footnotesize
  \caption{TOI-1710 parameters from the transit and RV joint fit, obtained with the model 1pGP. \label{table - JF POSTERIORS}}  
  \centering
  \begin{tabular}{lcc}
  \hline
  \hline
  Parameter & Prior$^{(\mathrm{a})}$ & Value$^{(\mathrm{b})}$   \\
  \hline
  \multicolumn{2}{l}{\emph{ \bf Model Parameters }} \\
    Orbital period $P_{\mathrm{orb}}$ [days]  & $\mathcal{U}[24.18, 24.38]$   & $ 24.283382 \pm 0.000019 $ \\
    Transit epoch $T_0$ [BJD - 2\,450\,000]  & $\mathcal{U}[9025.00, 9037.00]$  & $ 9031.23007_{-0.00037}^{+0.00041} $\\
    $\sqrt{e} \sin \omega_\star$ &  $\mathcal{U}(-1,1)$ & $0.42 \pm 0.13  $ \\
    $\sqrt{e} \cos \omega_\star$  &  $\mathcal{U}(-1,1)$ & $ -0.01 \pm 0.10$ \\
    Scaled planetary radius $R_\mathrm{p}/R_{\star}$ &  $\mathcal{U}[0,0.5]$ & $0.048 \pm 0.0004$   \\
    Impact parameter, $b$ &  $\mathcal{U}[0,1]$  & $ 0.081 _{-0.058} ^ {+0.093} $ \\
    Radial velocity semi-amplitude variation $K$ [m s$^{-1}$] &  $\mathcal{U}[0,50]$ & $ 4.20 \pm 1.0 $  \\ 
    \hline
    \multicolumn{2}{l}{\textbf{Derived parameters}} \\
    Planet radius [$R_{\rm J}$]  & $\cdots$ & $ 0.460 \pm 0.010 $ \\
    Planet radius [$R_{\oplus}$]  & $\cdots$ & $ 5.15 \pm 0.12 $  \\
    Planet mass [$M_{\rm J}$]  & $\cdots$ & $0.058_{-0.014} ^ {+0.015}  $    \\
    Planet mass [$M_{\oplus}$]  & $\cdots$ & $18.4_{-4.5} ^ {+4.8} $    \\
    Eccentricity $ecc$  & $\cdots$ & $0.185_{-0.091} ^ {+0.12}$   \\
    Scaled semi-major axis $a/R_\star$   & $\cdots$ & $ 30.6_{-3.9} ^ {+3.1}$  \\
    Semi-major axis $a$ [AU]  & $\cdots$ & $ 0.164 \pm 0.004 $  \\
    $\omega_{\rm P} $ (deg)  &  $\cdots$ &  91 $\pm$ 14 \\
    Orbital inclination $i$ (deg)  & $\cdots$ & $89.8 \pm 0.3 $  \\
    Transit duration $T_{14}$ [days] & $\cdots$ & $ 0.263_{-0.024} ^ {+0.037}$ \\
    Transit duration $T_{23}$ [days] & $\cdots$ & $ 0.238_{-0.021} ^ {+0.033} $ \\

    Equilibrium temperature$^{(\mathrm{c})}$ $T_{\rm eq}$ [K] & $\cdots$ & $ 730_{-30} ^ {+36} $ \\

    Planet instellation ${S}$ [$\mathrm{S}_\oplus$] & $\cdots$ & $ 33 \pm 2 $ \\
    
     \hline
    \multicolumn{2}{l}{\emph{ \bf Other system parameters}} \\
    Jitter term $\sigma_{\rm HARPS-N}$ [m s$^{-1}$] & $\mathcal{U}[0,60]$ & $1.40_{-0.33} ^{+0.35}$  \\
    Jitter term $\sigma_{\rm SOPHIE}$ [m s$^{-1}$] & $\mathcal{U}[0,60]$ & $4.4_{-1.2} ^{+1.3}$  \\
    Stellar density $\rho_\star$ [$\rho_{\odot}$] &  --  & $0.65 \pm 0.22$  \\
    Limb darkening $q_1$  & $\mathcal{N}[0.226,0.090]$ & $0.293 \pm 0.062 $  \\
    Limb darkening $q_2$ & $\mathcal{N}[0.52,0.21]$ & $0.34 \pm 0.11 $  \\
    \multicolumn{3}{l}{\emph{ \bf Stellar activity GP model Parameters}} \\
    $h_{\rm HARPS-N}$  [m s$^{-1}$]  &  $\mathcal{U}[0, 100]$ & $6.5_{-1.1}^{+1.4}$ \\
    $h_{\rm SOPHIE}$  [m s$^{-1}$]  &  $\mathcal{U}[0, 100]$ & $4.4_{-2.3}^{+2.4}$ \\
    $\lambda$  [days]  &  $\mathcal{U}[5, 2000]$ & $62 _{-12}^{+15}$ \\
    $\omega$    &  $\mathcal{U}[0.01, 1.50]$ & $0.375 _{-0.088}^{+0.110}$ \\
    $\theta$ (P$_{\rm rot}$)  [days]  &  $\mathcal{U}[2, 40]$ & $ 21.51_{-0.24}^{+0.23}$ \\
  \hline
   \noalign{\smallskip}
  \end{tabular}
~\\
  \emph{Note} -- $^{(\mathrm{a})}$ $\mathcal{U}[a,b]$ refers to uniform priors between $a$ and $b$; $\mathcal{N}[a,b]$ refers to Gaussian priors with median $a$ and standard deviation $b$.\\  
  $^{(\mathrm{b})}$ Parameter estimates and corresponding uncertainties are defined as the median and the 16th and 84th percentiles of the posterior distributions.\\
  $^{(\mathrm{c})}$ Assuming zero Bond albedo.\\
\end{table*}

We performed a joint fit using only \textit{TESS} photometry and RV data using {\tt PyORBIT}\footnote{Available at \url{https://github.com/LucaMalavolta/PyORBIT}} (\citealt{Malavoltaetal2016,Malavoltaetal2018}) in order to obtain the planetary system parameters. 
For the fit of transit light curves, this code makes use of the public package \texttt{batman} \citep{batman} and allows  Gaussian processes (GPs) to be included through the \texttt{george} package \citep{george} in order to model the presence of correlated noise or systematic effects in the data. The parameter space is sampled by using the MCMC technique with the ensemble sampler \texttt{emcee} \citep{foremanmackey2013}. The initial conditions are obtained with the global optimization code {\tt PyDE}.\footnote{Available at \url{https://github.com/hpparvi/PyDE}} We use the GP quasi-periodic kernel, as defined by \cite{Grunblattetal2015}, imposing a uniform prior on the rotational period $P_{\rm rot}$.  To run the GPs, we used 10\,000 steps, 184 walkers, a burn-in cut of 1\,000, and a thinning factor of 100. The confidence intervals of the posterior distributions are computed considering the 34.135th percentile from the median.

Regarding the photometric \textit{TESS} light curves, we modelled the time of first transit $T_c$, the orbital period $P$, the limb darkening (LD) with  \cite{Kipping2013} parametrization, the impact parameter $b$, and  the scaled planetary radius $R_{P}/R_{\star}$. The eccentricity $ecc$ and argument of periastron $\omega$ were modelled with the parametrization from \citealt{Eastmanetal2013} ($\sqrt{e}\cos\omega$,$\sqrt{e}\sin\omega$). We used Gaussian priors both for the stellar mass and radius (as obtained in Sect.\,\ref{subsec: Params}) and for the LD coefficients (calculated with {\tt PyLDTk};\footnote{Available at \url{https://github.com/hpparvi/ldtk}} \citealt{Parviainen2015, Husser2013}). The stellar density and the impact parameter $b$ were left free. In order to save computational time, we only used the \textit{TESS} data around TOI-1710\,b transits.

The RV data are modelled taking into account the offset between different instruments, and a jitter term for systematics and stellar noise. The joint fit was performed for three models: \textit{i)} one planet without GP (1p); \textit{ii)} one planet with GP (1pGP); \textit{iii)} one planet with GP and FWHM included in the GP (1pGP+FWHM).
This last model, where the GP is fed with the FWHM values, was tested because of the strong correlation between the FWHM and the HARPS-N RVs (Fig. \ref{Fig: correlation}). We computed the Bayesian criteria in order to evaluate their significance. In particular, we computed the  Bayesian information criterion (BIC) and the  Akaike information  criterion (AIC); 
the 1pGP+FWHM model is strongly preferred over the other two models (see Table \ref{tab:bicaic}), indicating that the activity plays an important role in the data.
This is also proved by the RV residuals, which are quite flat with no extra significant signals (FAP$\gg$10\%), when adopting the GP in the model (see Fig\,\ref{fig: GLS residuals}).
The priors, posteriors, and derived planetary parameters from the 1pGP+FWHM model are presented in Table\,\ref{table - JF POSTERIORS}. The phase-folded \textit{TESS} photometry and RVs are shown in Fig.\,\ref{fig:joint_fit_plot}.

\citet{Konig_TOI1710} also detected the $P_{\rm rot}$ in their activity indicators GLS periodograms, but the stellar rotation was not strong in their RV periodogram. No correlation plots (e.g. Fig.\,\ref{Fig: correlation}) are shown. They tested three RV models with different GP kernels  (one of which is the quasi-periodic kernel used in our analysis), and found values and uncertainties in good agreement with those obtained when no GP was included. However, they modelled the RV data with the \texttt{juliet} code (\citealp{juliet}), whose GPs can only be fed by the time stamps, as in our 1pGP model. The 1p and 1pGP models have no clear Bayesian criterion preference, but 1pGP+FWHM does (Table\,\ref{tab:bicaic}).
Although the $P_{\rm rot}$ and the planet period peaks are clear in the GLS periodogram (Fig.\,\ref{fig: GLS RVs}), they are close. Thus, it is expected that, when modelling the two models simultaneously, the semi-amplitude is smaller than the one obtained when only the planet Keplerian is considered (as in \citealp{Konig_TOI1710}). The semi-amplitudes obtained in this work and by \citet{Konig_TOI1710} differ by 1.6$\sigma$. Moreover, although our analyses included 54 more RVs, the fitted semi-amplitude has the same uncertainties ($\pm$1.0\,m\,s$^{-1}$) as that of \citet{Konig_TOI1710}. This may be explained by the addition of the GP in the model, and  in fact the   semi-amplitude uncertainty obtained with our 1p model is 0.8\,m\,s$^{-1}$.

By adding two more \textit{TESS} transits in the joint fit analysis, we improved the photometric parameters. We reduced the $P$, $T_0$, and $R_{\rm p}/R_{\star}$ uncertainties from 4.3$\cdot$10$^{-5}$, 4.2$\cdot$10$^{-4}$, and 6$\cdot$10$^{-4}$ in \citet{Konig_TOI1710} to 1.9$\cdot$10$^{-5}$, 3.7$\cdot$10$^{-4}$, and 4$\cdot$10$^{-4}$, respectively. However, the uncertainty for $R_{\rm p}$ did not improve (this work: 0.12\,$\mathrm{R}_\oplus$; \cite{Konig_TOI1710}: 0.11\,$\mathrm{R}_\oplus$), meaning that the main source of uncertainty comes from $R_{\star}$.

\section{Discussion and conclusions}
\label{Sec: Dicussion}

From our analyses, we derive for the transiting sub-Saturn planet TOI-1710\,b a radius of $R_{\rm p }$\,$=$\,5.15\,$\pm$\,0.12\,$\mathrm{R}_\oplus$, a mass of $M_{\rm p}$\,$=$\,18.4\,$^{+4.8}_{-4.5}$\,$\mathrm{M}_\oplus$, and a mean bulk density of $\rho_{\rm p}$\,$=$\,0.73\,$\pm$\,0.18\,$\mathrm{g\,cm^{-3}}$,
adopting the stellar parameter values from Section\,\ref{subsec: Params}.
Our planetary parameters are consistent with those previously derived by \cite{Konig_TOI1710}, and we confirmed the eccentric orbit of TOI-1710\,b (\textit{ecc}\,=\,0.185$_{-0.091} ^{+0.12}$). The mass uncertainties are very similar between the two works, but our mean mass measurement value presents a precision at the 4$\sigma$ level, and it is $\sim$1.5$\sigma$ away from that of \citet[$M_{\rm p}$\,$=$\,28.3\,$\pm$\,4.7\,$\mathrm{M}_\oplus$]{Konig_TOI1710}.
Although this difference is not significant, we needed to add a GP component accounting for the stellar rotation to successfully fit the RV dataset from this work (see Table\,\ref{tab:bicaic} and Fig.\,\ref{fig: GLS residuals}).


After subtracting the 1pGP+FWHM model, the GLS periodograms of the HARPS-N, SOPHIE, and combined RVs are well below the 10\,\% FAP (see Fig.\,\ref{fig: GLS residuals}). \cite{Konig_TOI1710} reported an extra peak at 30\,d with FAP\,$\simeq$\,2.2\,\% after modelling the  TOI-1710\,b signal. However, here with more RV measurements and after accounting for the stellar rotation, we do not find evidence of any extra signal in the  TOI-1710 planetary system.

\subsection{Potential atmospheric characterization of TOI-1710\,b}
\label{subsec: atmospheric characteriztion}

TOI-1710\,b is a gaseous planet with a scale height ($H$) of $\sim$390\,km.
As it is expected to posses a   significant envelope, we evaluated its suitability for atmospheric characterization via the transmission spectroscopy metric (TSM) proposed by \citet{Kempton2018_TSM}.
TSM takes into account the planet radius, mass,   equilibrium temperature,   stellar radius, and J magnitude.
The formula also includes  a scale factor to account for 10\,hours of observing time with the James Webb Space Telescope (JWST; \citealp{JWST_2006}) Near Infrared Imager and Slitless Spectrograph (NIRISS) instrument.
According to Table\,1 from \citet{Kempton2018_TSM}, the scale factor for sub-Saturn planets is 1.15.
The computed TSM for TOI-1710\,b is $\sim$150, which is greater than the threshold TSM of 90 for planets ranging 1.5--10\,$\mathrm{R}_\oplus$ to be selected as high-quality targets for its atmospheric characterization.

There are only a few sub-Saturn planets around bright stars (J\,$\leq$\,9) with higher TSM: GJ\,436b (J\,=\,6.9\,mag, TSM\,=\,481; \citealp{GJ436b}), HAT-P-11b (J\,=\,7.6\,mag, TSM\,=\,202; \citealp{HAT-P-11b_discovery}), GJ\,3470b (J\,=\,8.8\,mag, TSM\,=\,270; \citealp{GJ3470b}), HD\,89345b (J\,=\,8.1\,mag, TSM\,=\,114; \citealp{HD89345b}), WASP-166b (J\,=\,8.4\,mag, TSM\,=\,233; \citealp{WASP-166b}), HD\,219666b (J\,=\,8.6\,mag, TSM\,=\,142; \citealp{HD219666b}), LTT\,9779b (J\,=\,8.4\,mag, TSM\,=\,185; \citealp{LTT9779b}), and TOI-421c (J\,=\,8.5\,mag, TSM\,=\,161; \citealp{TOI421c}).
Among the many species that are expected in TOI-1710\,b's atmosphere, one of the most interesting and accessible would be the \ion{He}{i} triplet, which has   proved to be a powerful tool for studying extended planetary atmospheres, tracking mass-loss and winds in the upper atmospheres, and even detecting cometary-like atmospheric tails \citep{Nortmann2018}. 
\ion{He}{i} has already been detected  in similar sub-Saturns, such as GJ\,3470\,b \citep{He_GJ3470b, Ninan2020ApJ...894...97N}, HAT-P-11\,b \citep{He_HAT-P-11b}, and HAT-P-26\,b \citep{HAT-P-26_He}.
We used the rotational period of the star to calculate the expected X-ray and extreme UV emission, using the relations in \citet{wri11} and \citet{san11}.
We estimate an X-ray luminosity of $\log L_{\rm X}$\,$=$\,28.07 and a calculated $F_{\rm XUV}$\,$\sim$1\,W\,m$^{-2}$ irradiance in the 5--504\,$\AA$ spectral range at the planet b orbit, similar to the flux received by HD\,209458\,b, where the \ion{He}{i} triplet was positively detected by \citet{alo19}.
This is further supported by the high estimated transmission spectroscopy S/N calculated by \citet{Konig_TOI1710}, placing TOI-1710\,b as the fourth-best candidate within the sub-Saturn population. It is thus clear that TOI-1710\,b is a very appealing target for the atmospheric studies, and strong atmospheric features (e.g. the \ion{Na}{i} doublet and \ion{K}{i} line at 7701\,$\AA$ in the visible or the \ion{He}{i} triplet and molecules in the near-infrared) are expected in TOI-1710\,b, if they are not veiled by clouds or hazes.

\begin{figure*}
    \centering
    \includegraphics[width=\textwidth]{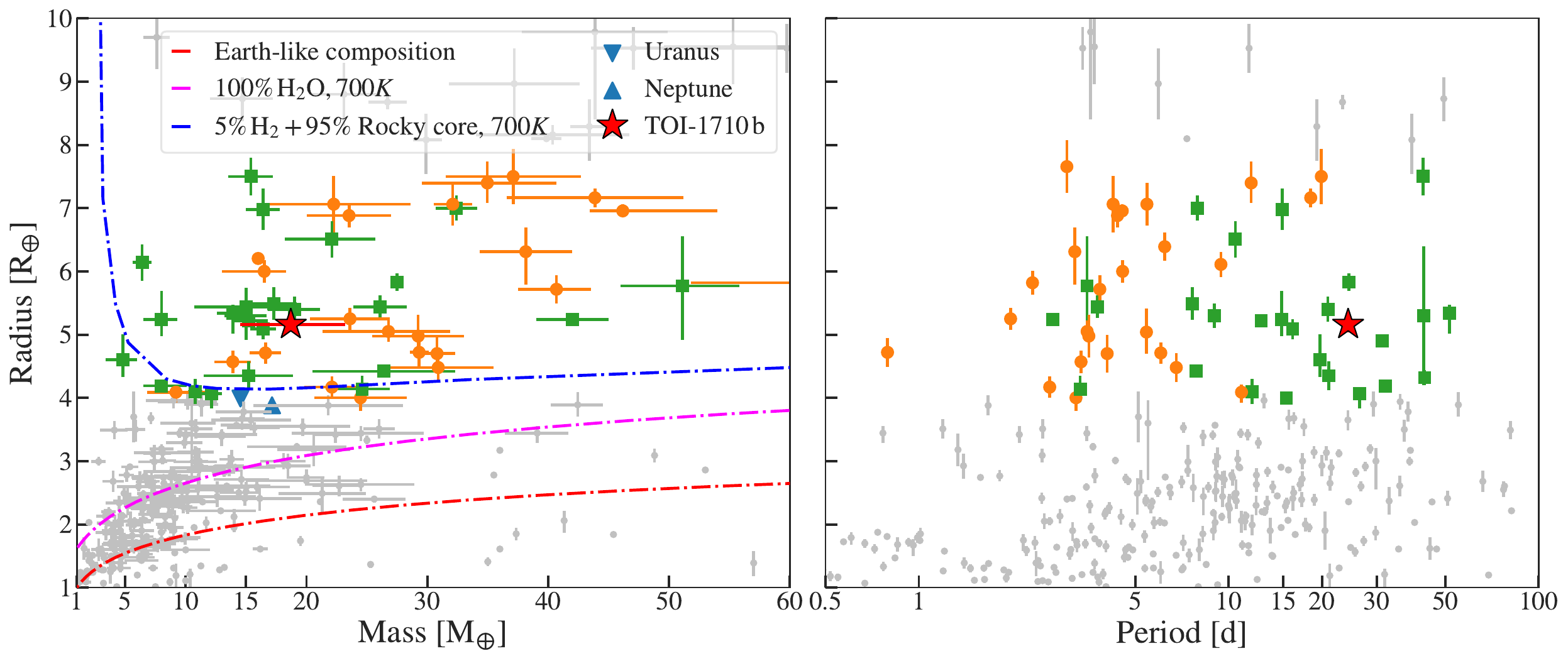}
    \caption{Mass-radius (\textit{left}) and period-radius (\textit{right}) diagrams for all known planets with $M_{\rm p}$\,$\in$\,[1,60]\,M$_{\oplus}$ and $R_{\rm p}$\,$\in$\,[1,10]\,R$_{\oplus}$ determined with a precision better than 30\% (grey points). The coloured dashed lines are the theoretical mass-radius models from \citet{Zeng_models}. Sub-Saturns located in multi-planet systems are marked with green squares, and sub-Saturns located in single-planet systems are orange circles. TOI-1710\,b is marked with a red star. }
    \label{Fig: Mass-Radius}
\end{figure*}

\begin{figure}
    \centering
    \includegraphics[width=\hsize]{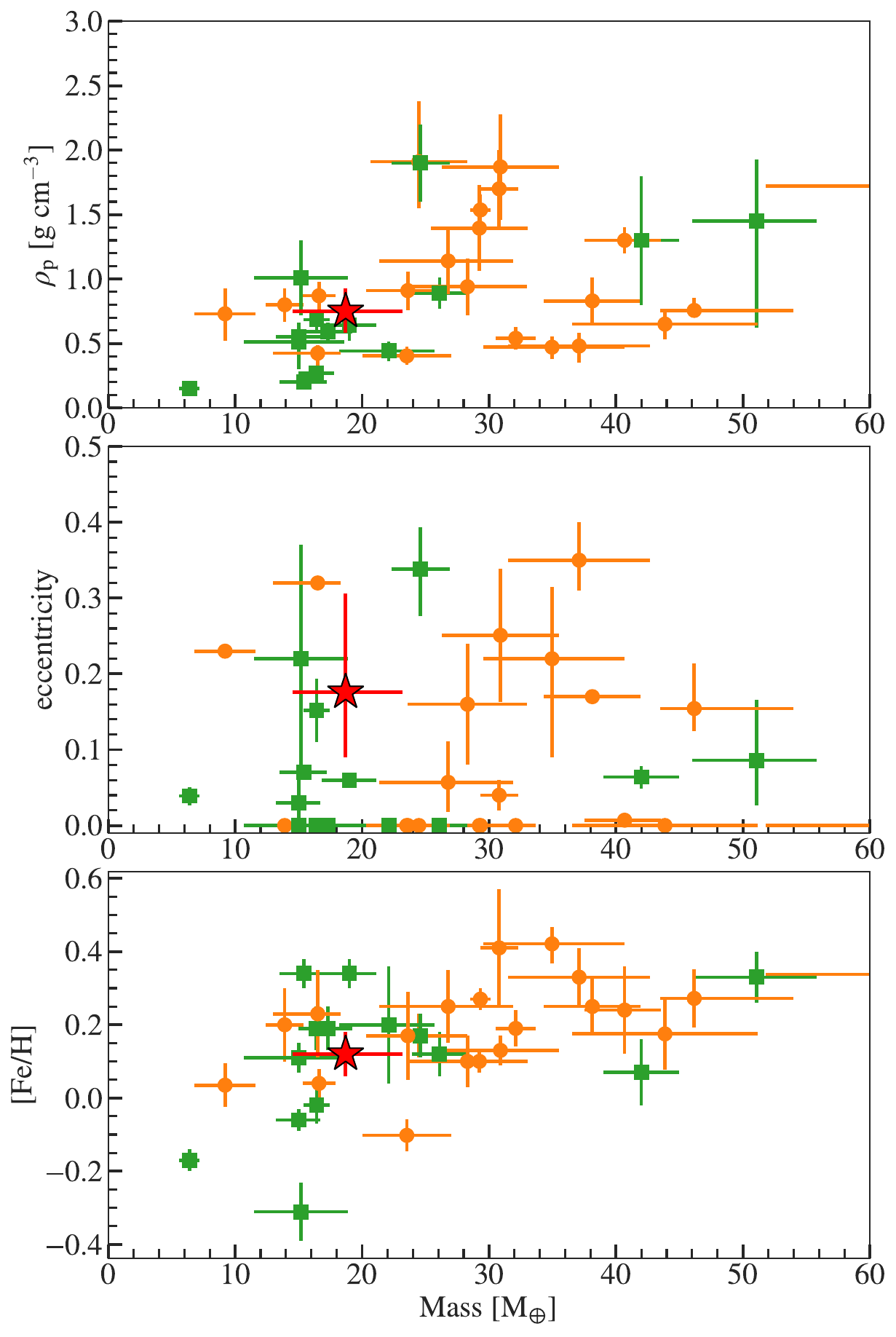}
    \caption{ Mass-density (\textit{top}), mass-eccentricity (\textit{middle}), and mass-metalicity (\textit{bottom}) diagrams for the sub-Saturn planets from Fig.\,\ref{Fig: Mass-Radius}. Sub-Saturns located in multi-planet systems are marked with green squares and sub-Saturns located in single-planet systems are orange circles. TOI-1710\,b is marked with a red star.
    }
    \label{Fig: Density-Ecc-Metal}
\end{figure}

\subsection{TOI-1710\,b within the sample of sub-Saturns}
\label{subsec: planet composition}

Figure\,\ref{Fig: Mass-Radius} compares TOI-1710\,b with known exoplanets, with masses and radii measured with a precision better than 30\%, from the NASA Exoplanet Archive\footnote{\url{ https://exoplanetarchive.ipac.caltech.edu/ }} along with theoretical composition models of \citet{Zeng_models}.\footnote{\url{ https://lweb.cfa.harvard.edu/~lzeng/planetmodels.html }}
The displayed models regard planets with Earth-like composition cores (32.5\%\,Fe\,+\,67.5\%\,MgSiO$_2$), 100\%\,H$_2$O composition, and 95\%\,Earth-like rocky core surrounded by 5\%\,H$_2$ gaseous envelopes.
The models are truncated at a reference pressure of 1\,mbar and consider an isothermal atmosphere, for which we chose an equilibrium temperature closest to that of  TOI-1710\,b.
Its position in the mass-radius diagram (Fig.\,\ref{Fig: Mass-Radius},\,left panel), well above the lighter planetary model available (Earth-like composition core with 5\%\,H$_2$ gaseous envelope), and its bulk mean density ($\rho_{\rm p}$\,$=$\,0.75\,$\pm$\,0.18\,$\mathrm{g\,cm^{-3}}$) suggest that TOI-1710\,b may hold a large H--He envelope that could be suitable for spectroscopical exploration,  due to the brightness of its host star.

In Figure\,\ref{Fig: Mass-Radius} (and in Figure\,\ref{Fig: Density-Ecc-Metal} as well), we marked all other known sub-Saturns detected in multi-planet systems as green squares, while sub-Saturns in single-planet systems (e.g. TOI-1710\,b) are shown as orange circles. 
Figure\,\ref{Fig: Mass-Radius},\,\textit{left panel}, shows that lone  sub-Saturns tend to be more massive, which is a trend already reported in previous studies (e.g. \citealp{Petigura2017}, \citealp{HD89345b}, and \citealp{Nowak_2020}). However, in the intermediate region at 20--30\,$\mathrm{M}_\oplus$, where TOI-1710\,b is located, the two populations are mixed, and there is not a clear boundary line between them. Its closer lone sub-Saturn companion in the mass-radius diagram is TOI-674\,b \citep{Murgas_TOI}, which has an orbital period of less than 2\,days ($P$\,=\,1.97\,d).

The right panel in Fig.\,\ref{Fig: Mass-Radius} displays the period-radius diagram, where the  TOI-1710\,b position seems to be an outlier of the single-planet sub-Saturn population.
From this figure we can clearly see that sub-Saturns in multi-planet systems tend to orbit host stars with longer orbital periods than those in single-planet systems. This is unlikely to be an observational or instrumental bias because the lighter planets are the ones farther from their stars.
However, because single sub-Saturns have short periods, their possible companions may be in outer orbits. Thus, the hypothetical companions are less likely to transit and should be massive enough to be detectable via radial velocity or astrometry methods.
TOI-1710\,b has a period of 24\,d, and  is well within the parameter space of the multi-planet systems;  it is the sub-Saturn planet in a single-planet system with the longest period in the diagram. However, there are other sub-Saturn planets with longer periods than TOI-1710\,b whose properties have not been determined with precision, for example Kepler-413A\,b \citep{Kepler-413b} with $P$\,=\,66\,d, and $M_{\rm p}$\,$=$\,67\,$\pm$\,21\,$\mathrm{M}_\oplus$.
From both panels of Fig.\,\ref{Fig: Mass-Radius}, we conclude that single sub-Saturns are more massive and have shorter periods than accompanied sub-Saturns, which are lighter and have longer periods.

\citet{Petigura2017} and \citet{Nowak_2020} also explored the possible relations between metallicity, density, and eccentricity of sub-Saturns with   planetary mass.
In Figure\,\ref{Fig: Density-Ecc-Metal} we reproduced these diagrams;   TOI-1710\,b is indicated  with a red star.
Other sub-Saturns are marked following the criteria in   Fig.\,\ref{Fig: Mass-Radius}.
Those works found a marginal correlation (Spearman correlation coefficient $r$\,=\,0.57) between the stellar metallicity and the mass of the sub-Saturns, where massive planets orbit metal-rich stars.
A slighly enriched metallicity of $\rm [Fe/H]$\,=\,0.12\,$\pm$\,0.06\,dex places TOI-1710 at the centre of the metallicity distribution (see Fig.\,\ref{Fig: Density-Ecc-Metal},\,\textit{bottom panel}).
According to the mass-metallicity correlation, the planet mass should also be   in the centre of the planetary masses, which TOI-1710\,b is.
TOI-1710\,b is displayed in central positions in all the panels of Fig.\,\ref{Fig: Density-Ecc-Metal}, and in the mass-radius diagram (Fig.\,\ref{Fig: Mass-Radius},\,\textit{left panel}). Thus, although we have not found evidence of extra planetary signals, further follow-up of the  TOI-1710 system will allow us to keep refining the characteristics of the transiting planet, and confirm if TOI-1710\,b is alone orbiting its host star or if it is the inner planet of a multi-planet system.



\begin{acknowledgements}
This work is partly financed by the Spanish Ministry of Economics and Competitiveness through grants PGC2018-098153-B-C31.

We acknowledge the contributions of Deven Combs, Sudhish Chmaladinne, Kevin Eastridge and Michael Bowen in the collection and anaylsis of the TOI 1710.01 ground-based light curve from George Mason Observatory.

G.N. thanks for the research funding from the Ministry of Education and Science programme the "Excellence Initiative - Research University" conducted at the Centre of Excellence in Astrophysics and Astrochemistry of the Nicolaus Copernicus University in Toru\'n, Poland.

T.M. acknowledges financial support from the Spanish Ministry of Science and 
Innovation (MICINN) through the Spanish State Research Agency, under the Severo Ochoa Program 2020-2023 (CEX2019-000920-S)

T.H. acknowledges support from the European Research Council under the Horizon 2020 Framework Program via the ERC Advanced Grant Origins 83 24 28.

P. D. is supported by a National Science Foundation (NSF) Astronomy and Astrophysics Postdoctoral Fellowship under award AST-1903811.

D. D. acknowledges support from the TESS Guest Investigator Program grant 80NSSC19K1727 and NASA Exoplanet Research Program grant 18-2XRP18-2-0136.

J.O.M. agraeix el recolzament, suport i \`anims que sempre ha rebut per part de padrina Conxa, padrina Merc\`e, Jeroni, Merc\`e i m\'es familiars i amics.
J.O.M acknowledges the special support from Maite, Guillem, Alejandro, Benet, Joan, and Montse. This research has made use of resources from AstroPiso collaboration. J.O.M. acknowledges the contributions of Jorge Terol Calvo, i molt especialment a tu, Yess.
Based on observations made with the Italian Telescopio Nazionale Galileo (TNG) operated on the island of La Palma by the Fundación Galileo Galilei of the INAF (Istituto Nazionale di Astrofisica) at the Spanish Observatorio del Roque de los Muchachos of the Instituto de Astrofisica de Canarias

This work has been carried out within the framework of the NCCR PlanetS supported by the Swiss National Science Foundation.
 
This paper includes data collected by the TESS mission, which are publicly available from the Mikulski Archive for Space Telescopes (MAST).
Funding for the \textit{TESS} mission is provided by NASA's Science Mission directorate.
We acknowledge the use of public TESS data from pipelines at the TESS Science Office and at the TESS Science Processing Operations Center.
Resources supporting this work were provided by the NASA High-End Computing (HEC) Program through the NASA Advanced Supercomputing (NAS) Division at Ames Research Center for the production of the SPOC data products.
This research has made use of the Exoplanet Follow-up Observation Program website, which is operated by the California Institute of Technology, under contract with the National Aeronautics and Space Administration under the Exoplanet Exploration Program.

This work has made use of data from the European Space Agency (ESA) mission {\it Gaia} (\url{https://www.cosmos.esa.int/gaia}), processed by the {\it Gaia} Data Processing and Analysis Consortium (DPAC,\url{https://www.cosmos.esa.int/web/gaia/dpac/consortium}). Funding for the DPAC has been provided by national institutions, in particular the institutions participating in the {\it Gaia} Multilateral Agreement.

Some of the observations in the paper made use of the High-Resolution Imaging instrument ‘Alopeke under Gemini LLP Proposal Number: GN/S-2021A-LP-105.‘Alopeke was funded by the NASA Exoplanet Exploration Program and built at the NASA Ames Research Center by Steve B. Howell, Nic Scott, Elliott P. Horch, and Emmett Quigley. ‘Alopeke  was mounted on the Gemini North telescope of the international Gemini Observatory, a program of NSF’s OIR Lab, which is managed by the Association of Universities for Research in Astronomy (AURA) under a cooperative agreement with the National Science Foundation. on behalf of the Gemini partnership: the National Science Foundation (United States), National Research Council (Canada), Agencia Nacional de Investigación y Desarrollo (Chile), Ministerio de Ciencia, Tecnología e Innovación (Argentina), Ministério da Ciência, Tecnologia, Inovações e Comunicações (Brazil), and Korea Astronomy and Space Science Institute (Republic of Korea).

This work made use of \texttt{tpfplotter} by J. Lillo-Box (publicly available in \url{www.github.com/jlillo/tpfplotter}), which also made use of the python packages \texttt{astropy}, \texttt{lightkurve}, \texttt{matplotlib} and \texttt{numpy}.
This work made use of \texttt{corner.py} by Daniel Foreman-Mackey (\citealp{CORNER_PLOT}).

\end{acknowledgements}

%
%

\bibliographystyle{aa}
\bibliography{references}


\begin{appendix}
\label{Sec:Appendix}

\section{Additional light curve  figures and tables}
\label{App: LC additional}

\begin{figure*}[h]
    \centering
    \includegraphics[width=\textwidth]{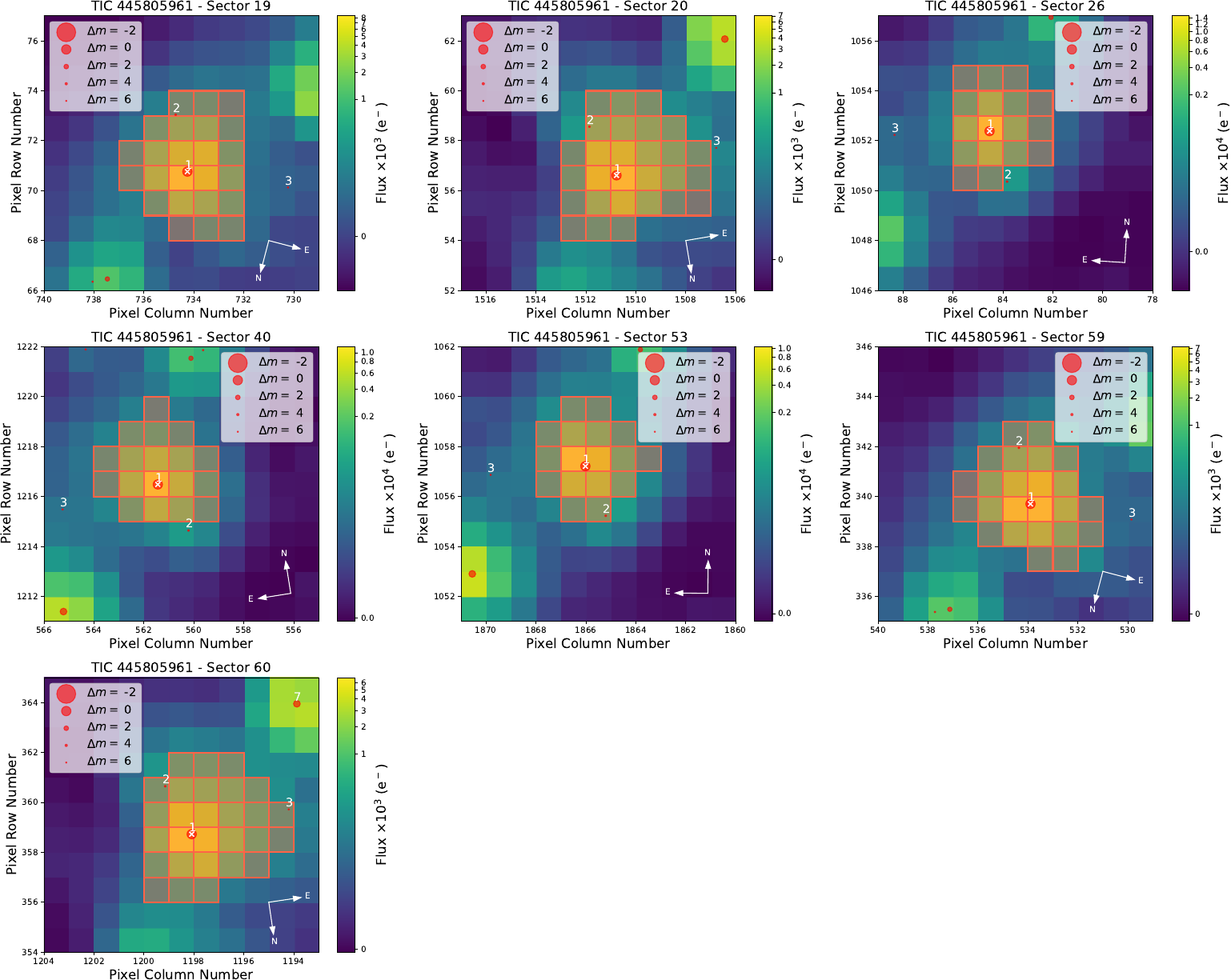}
    \caption{\textit{TESS} target pixel file image of TOI-1710 (TIC 445805961) observed in Sectors\,19, 20, 26, 40, 53, 59, and 60 (made with \texttt{tpfplotter}). The pixels highlighted in red show the aperture used by \textit{TESS} to obtain the photometry. The electron counts are colour-coded. The positions and sizes of the red circles respectively represent the positions and \textit{TESS} magnitudes of nearby stars. TOI-1710 is marked with a cross ($\times$) and labelled  $\#$1. \label{Fig: TESS_TPF_S19_S20_S26}}
\end{figure*}

\begin{table}[h]
\caption[width=\columnwidth]{
\label{table - juliet priors and posteriors} Prior and posterior distributions from the \texttt{juliet} photometric fit. Prior labels $\mathcal{U}$, $\mathcal{N}$, $\mathcal{F}$, and $\mathcal{J}$ represent uniform, normal, fixed, and Jeffrey's distribution, respectively.
}
\centering
\resizebox{\columnwidth}{!}{%
\begin{tabular}{lcc}

\hline \hline 
\noalign{\smallskip} 

Parameter & Prior & Posterior \vspace{0.05cm}\\
\hline
\noalign{\smallskip}

$P$ [d] & $\mathcal{N}(24.28,0.01)$ & 24.283385\,(18)  \vspace{0.05cm} \\ 
$t_0$\,$^{(a)}$ & $\mathcal{N}(1836.96,0.01)$ & 1836.96299\,(48)  \vspace{0.05cm} \\ 
\textit{ecc} (deg) & $\mathcal{F }(0)$ & --  \vspace{0.05cm} \\ 
$\omega$ (deg) & $\mathcal{F }( 90)$ & --  \vspace{0.05cm} \\ 
$r_{1}$ & $\mathcal{U }(0,1 )$ & 0.409$^{+0.059}_{-0.052}$  \vspace{0.05cm} \\ 
$r_{2}$ & $\mathcal{ U}(0,1 )$ & 0.04990$\pm$0.00037  \vspace{0.05cm} \\ 
$\rho_{\star}$ [kg\,m$^{-3}$]  & $\mathcal{ N }(1620,100)$ & 1617$^{+36}_{-64}$  \vspace{0.05cm} \\
$\mu_{\textit{TESS}}$ (ppm) & $\mathcal{N }(0.0,0.1)$ & 135$\pm$6  \vspace{0.05cm} \\ 
$\sigma_{\textit{TESS}}$ (ppm) & $\mathcal{J }( 10^{-6}, 10^{6})$ & 225$\pm$20  \vspace{0.05cm} \\ 
$q_{1,\textit{TESS}}$ & $\mathcal{U }(0,1 )$ & 0.23$^{+0.10}_{-0.07}$  \vspace{0.05cm} \\ 
$q_{2,\textit{TESS}}$ & $\mathcal{ U}(0,1 )$ & 0.52$^{+0.21}_{-0.17}$  \vspace{0.05cm} \\ 
GP$_\mathrm{\sigma}$ (ppm) & $\mathcal{J }(10^{-6}, 10^{6})$ & 224$\pm$16 \vspace{0.05cm} \\ 
GP$_\mathrm{\rho}$ [d]  & $\mathcal{J }(10^{-3}, 10^{3})$ & 0.93$^{+0.08}_{-0.07}$  \vspace{0.05cm} \\

\noalign{\smallskip}
\hline
\end{tabular}
}

\tablefoot{ $^{(a)}$ Central time of transit ($t_0$) units are BJD\,$-$\,2\,457\,000.}
\end{table}

\begin{figure*}[h]
    \centering
    \includegraphics[width=\textwidth]{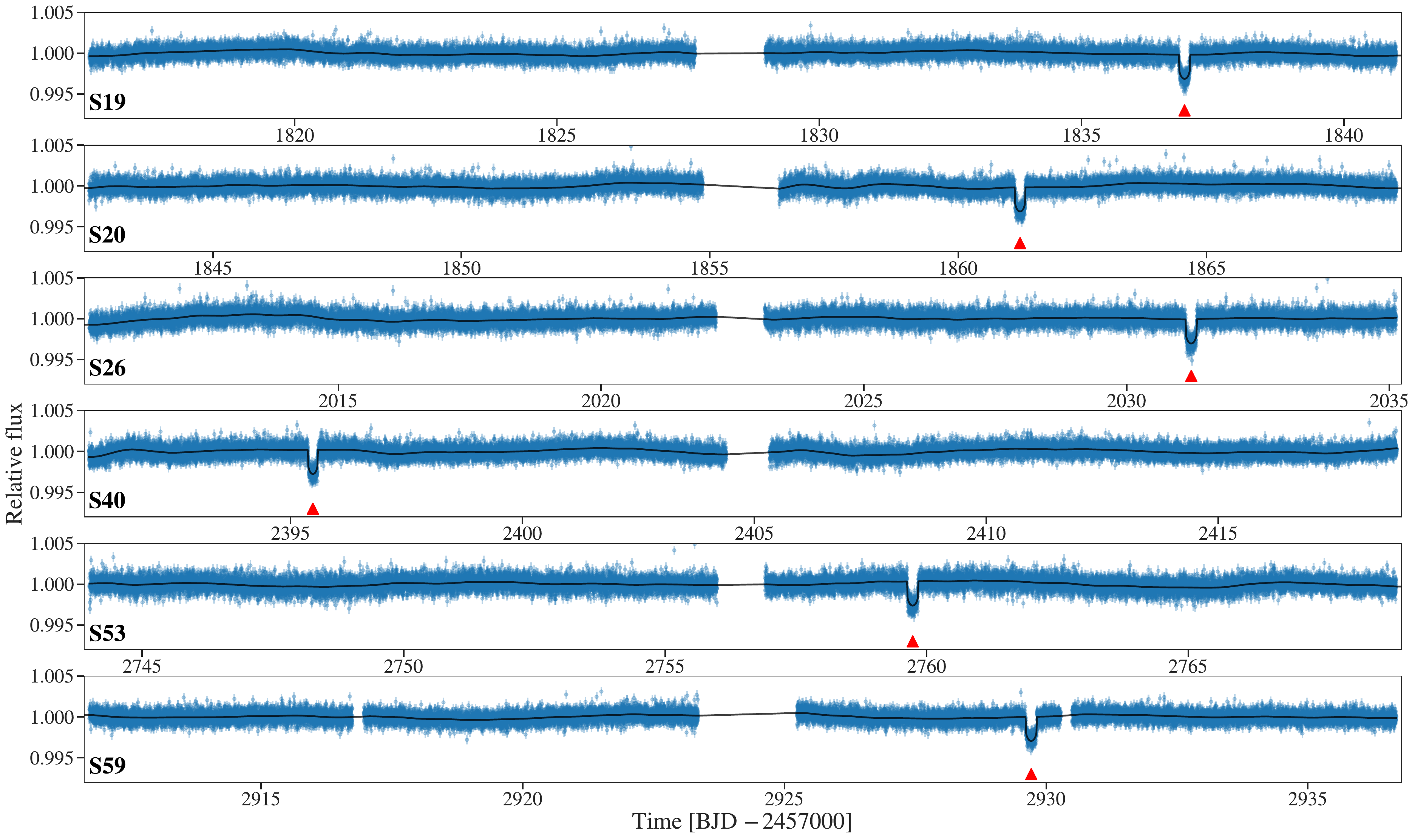}
    \caption{TOI-1710 2 min cadence light curve from \textit{TESS} Sectors\,19, 20, 26, 40, 53, and 59 (blue points with error bars) along with the transit plus GP model (black line) from \texttt{juliet}. Upward-pointing red triangles mark the  TOI-1710\,b transits.}
    \label{Fig: juliet TESS SECTORS}
\end{figure*}

\begin{figure}
    \centering
    \includegraphics[width=\hsize]{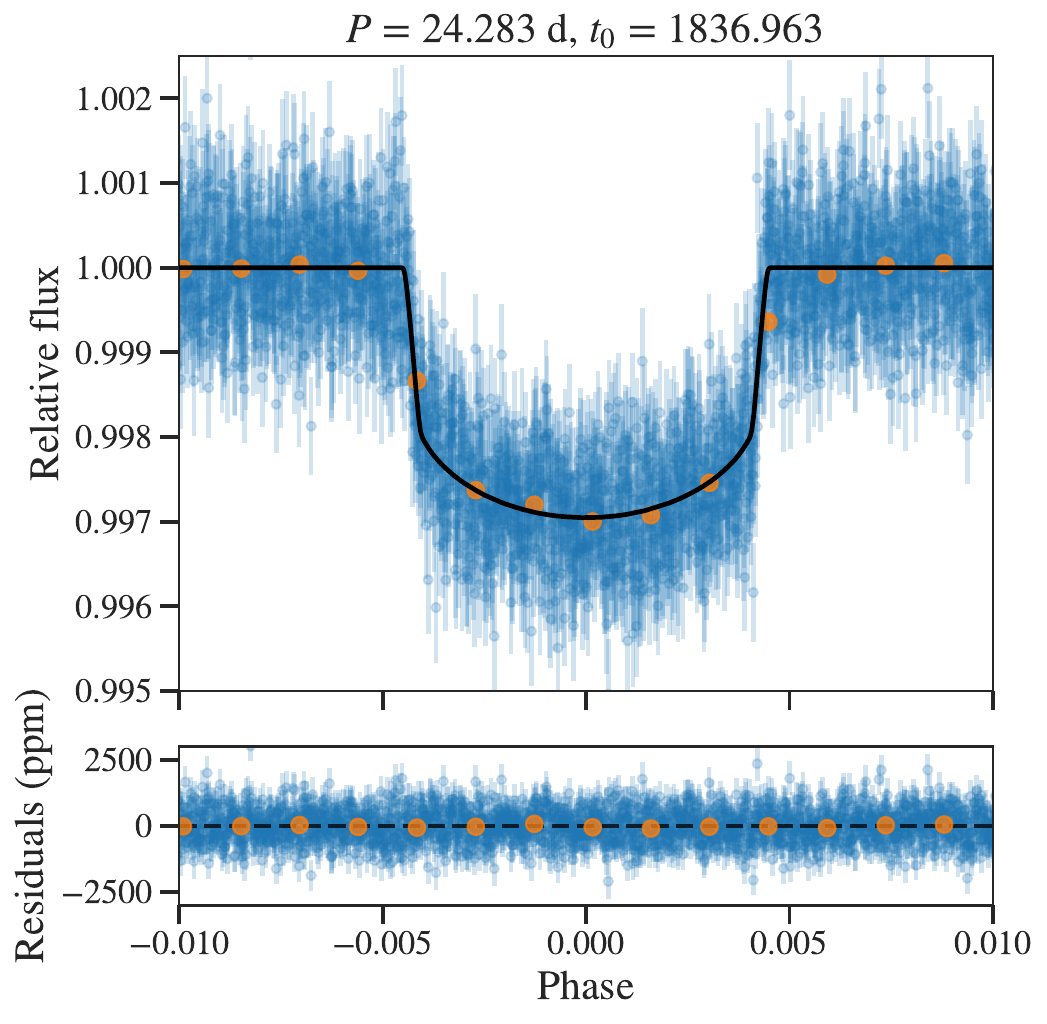}
    \caption{TOI-1710 \textit{TESS} photometry (blue dots with error bars) phase-folded to the period $P$ and central time of transit $t_0$ (shown above each panel, $t_0$ units are BJD\,$-$\,2\,457\,000) derived from the \texttt{juliet} fit. The black line is the best transit model for TOI-1710\,b. The orange points show the  binned photometry for ease of visualization. The GP model was removed from the data.}
    \label{Fig: juliet PHASE FOLDED}
\end{figure}

\begin{figure}
    \centering
    \includegraphics[width=\hsize]{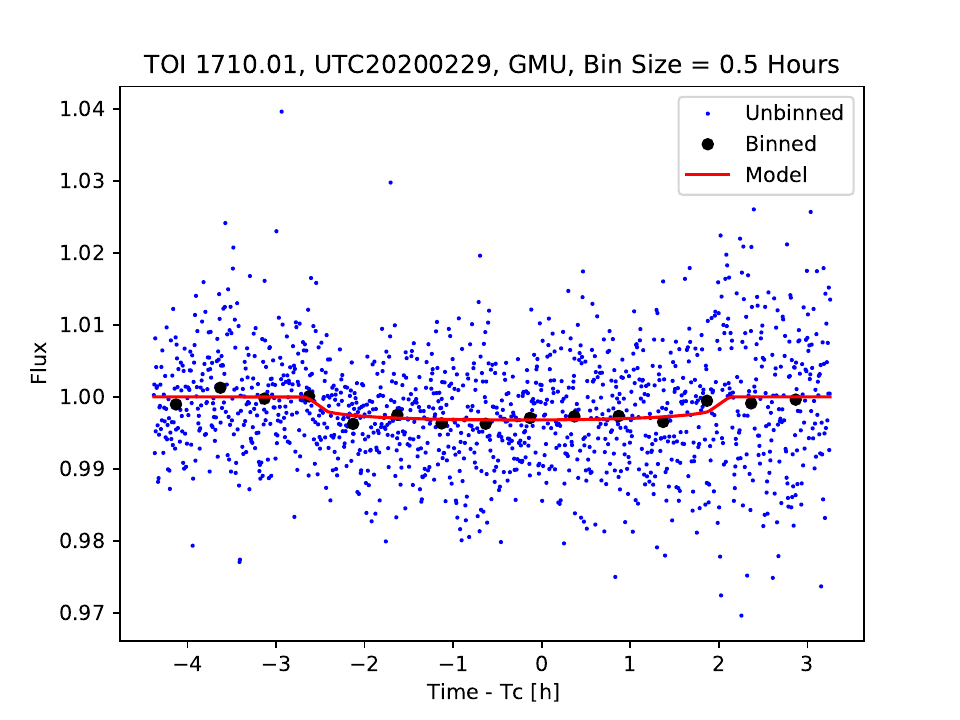}
    \caption{Light curve from the  George Mason University Observatory's 0.8 m telescope along with the best-fit transit model (red line). The blue points show the original photometric data with a cadence of 20\,s; the black points show 30 min binned photometry.}
    \label{Fig: GB Photometry}
\end{figure}

\clearpage

\section{Additional radial velocity  figures and tables}
\label{App: Joint Fit additional}


\input{Table_RVs_1}
\input{Table_ACT_1}

\input{Table_RVs_2}
\input{Table_ACT_2}

\begin{figure}[h]
    \centering
    \includegraphics[width=0.49\textwidth]{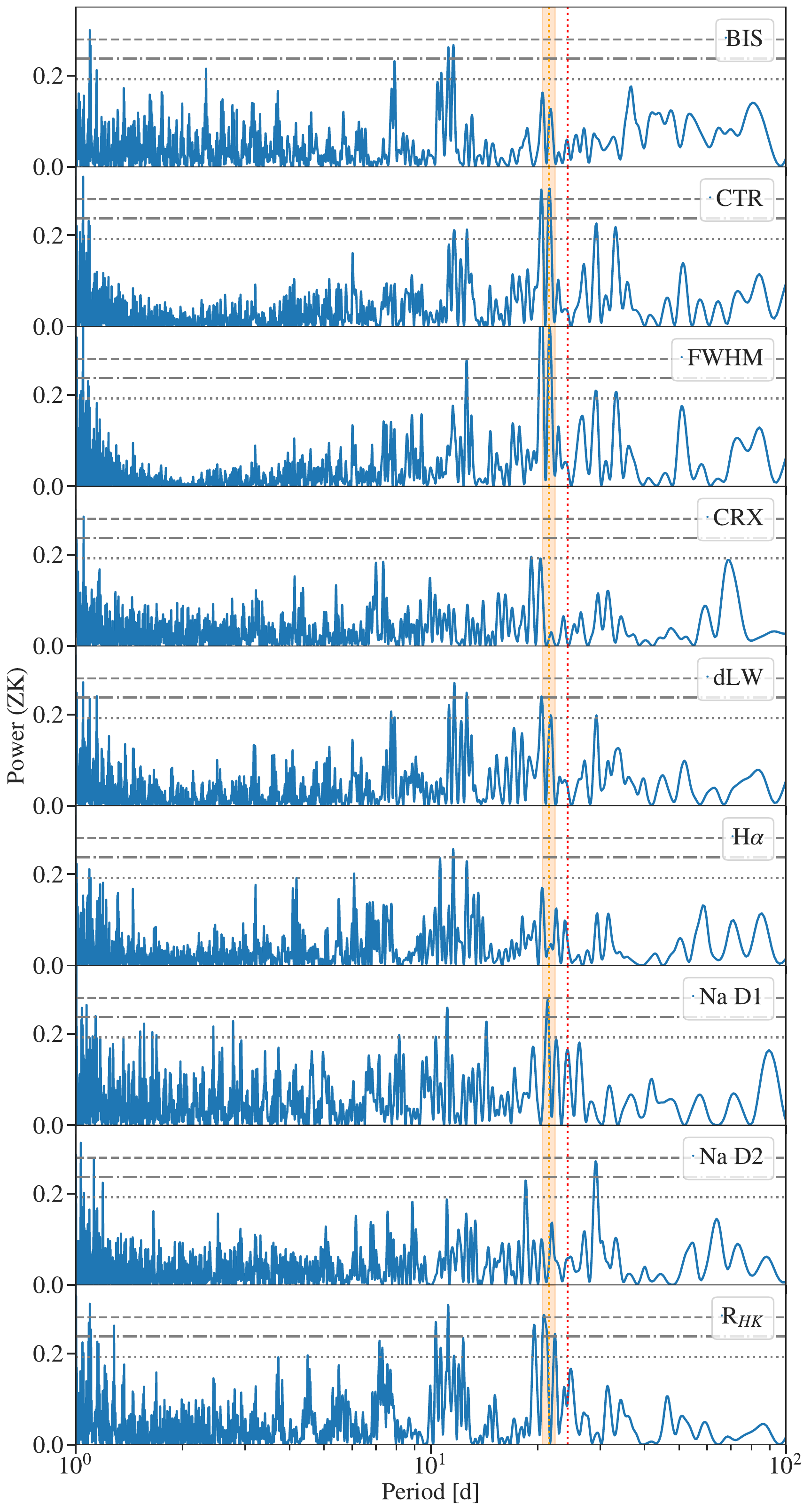}
    \caption{GLS periodograms for the activity indicators computed from the HARPS-N spectra (from top to bottom): cross-correlation function bisector (BIS), cross-correlation function contrast (CTR), cross-correlation function full width at half maximum (FWHM), chromatic index (CRX), differential line width (dLW), and the  H$\alpha$, Na\,D1, Na\,D2, and R$_{\rm HK}$ indexes. The 10\%, 1\%, and 0.1\% FAP levels are indicated by the grey dotted, dash-dotted, and  dashed lines, respectively.
    The vertical red dotted line indicates the period of TOI-1710\,b at 24.28\,d as a reference. The shaded orange region indicates the 3$\sigma$ region of the $P_{\rm rot}$ from the joint fit.
    }
    \label{Fig:ACT}
\end{figure}

\begin{figure*}
    \centering
    \includegraphics[width=\textwidth]{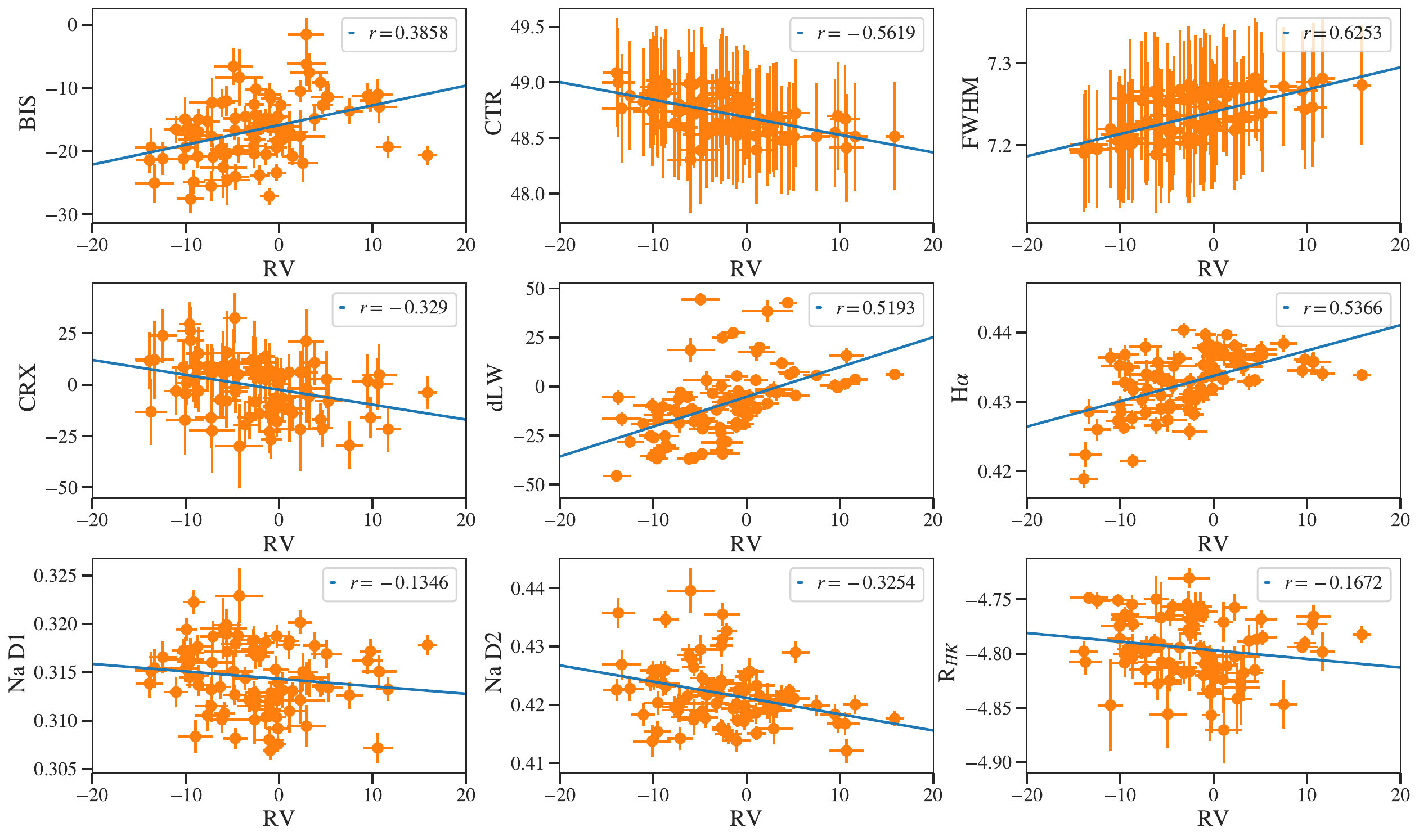}
    \caption{Correlation between HARPS-N RVs and the HARPS-N activity indicators (orange points with error bars). In each case a linear regression was performed (overplotted blue line) and  the Pearson correlation $r$ was computed, which is indicated.}
    \label{Fig: correlation}
\end{figure*}

\begin{figure*}
  \centering
  \includegraphics[width=0.33\textwidth]{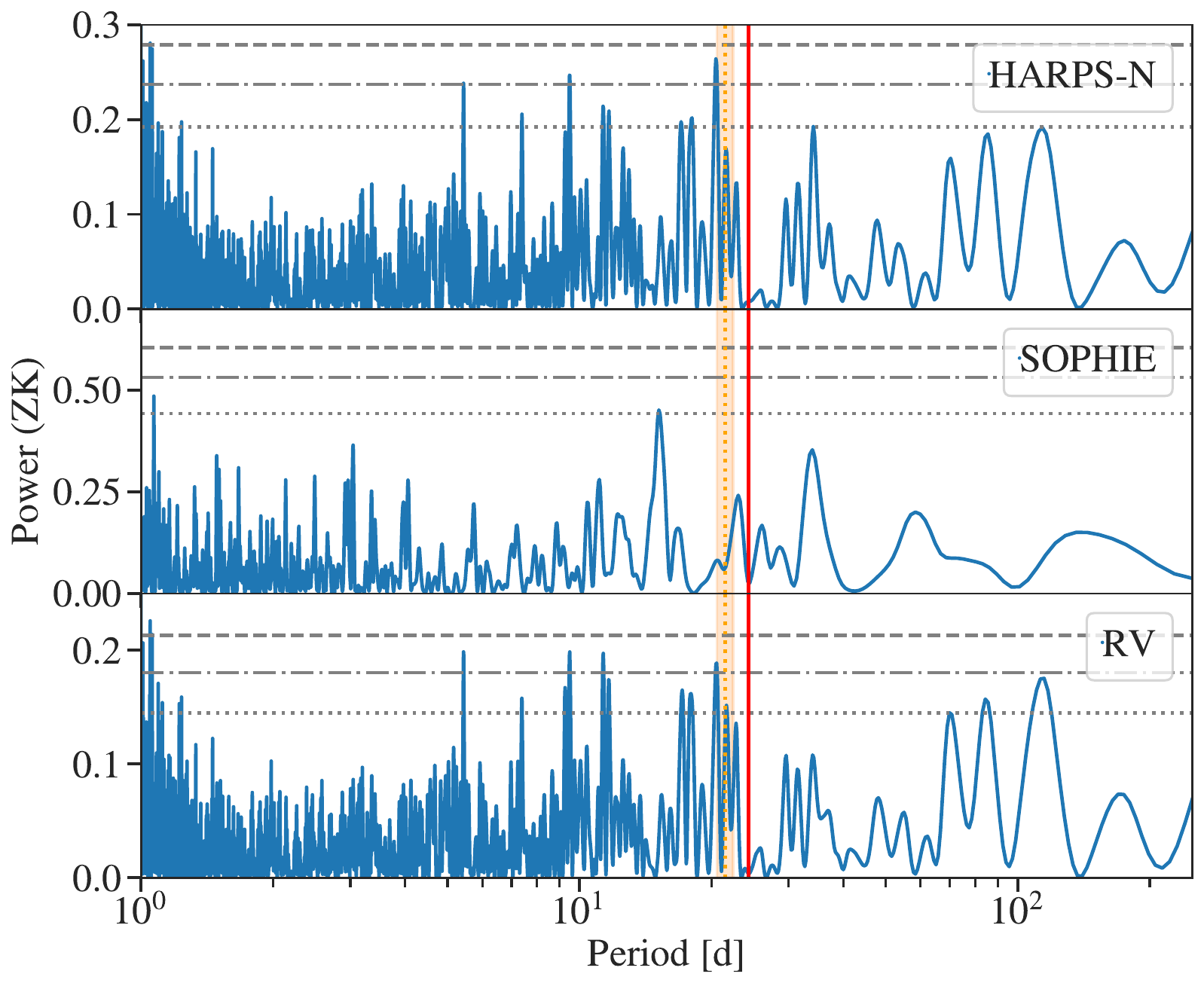}
  \includegraphics[width=0.33\textwidth]{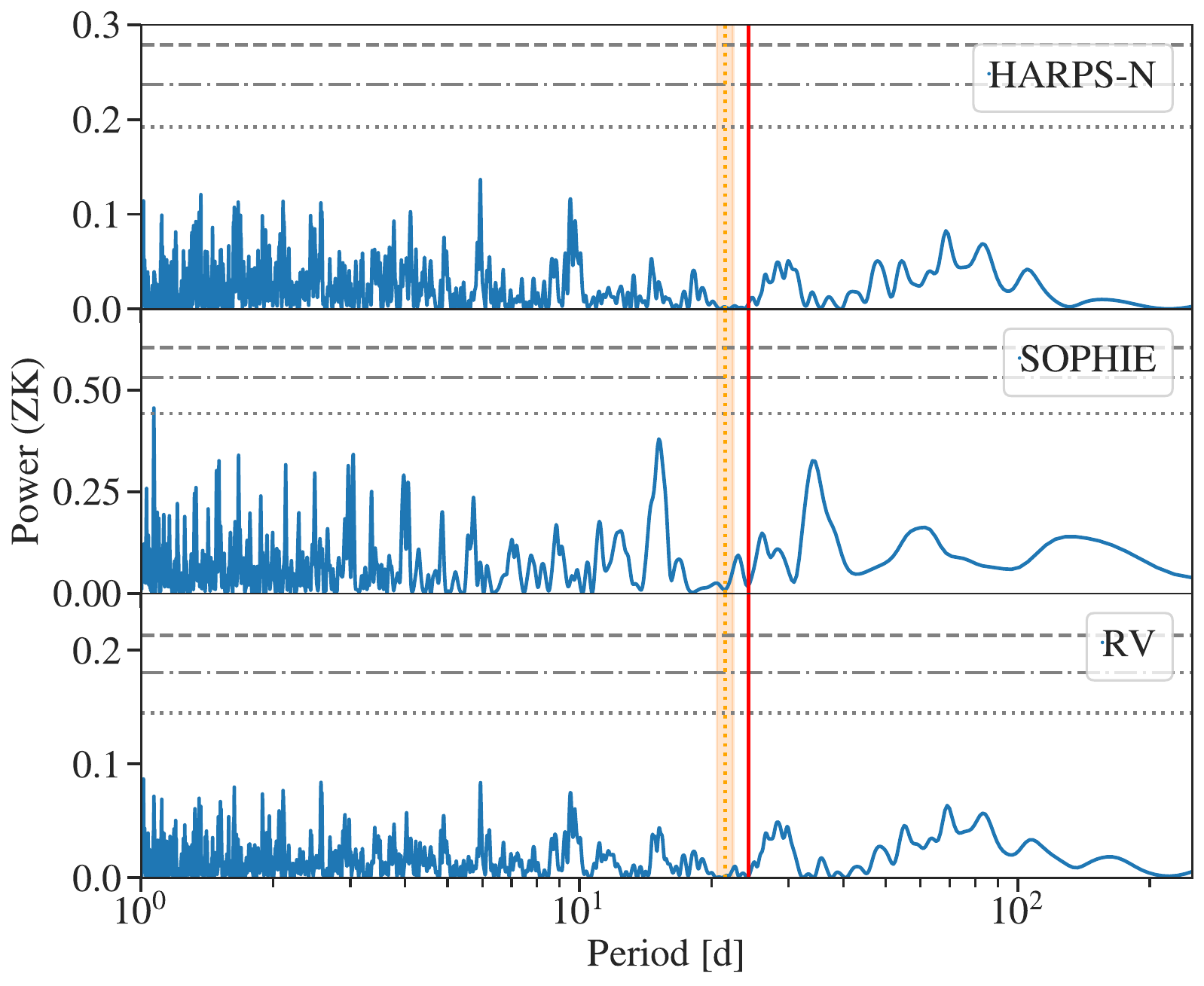}
  \includegraphics[width=0.33\textwidth]{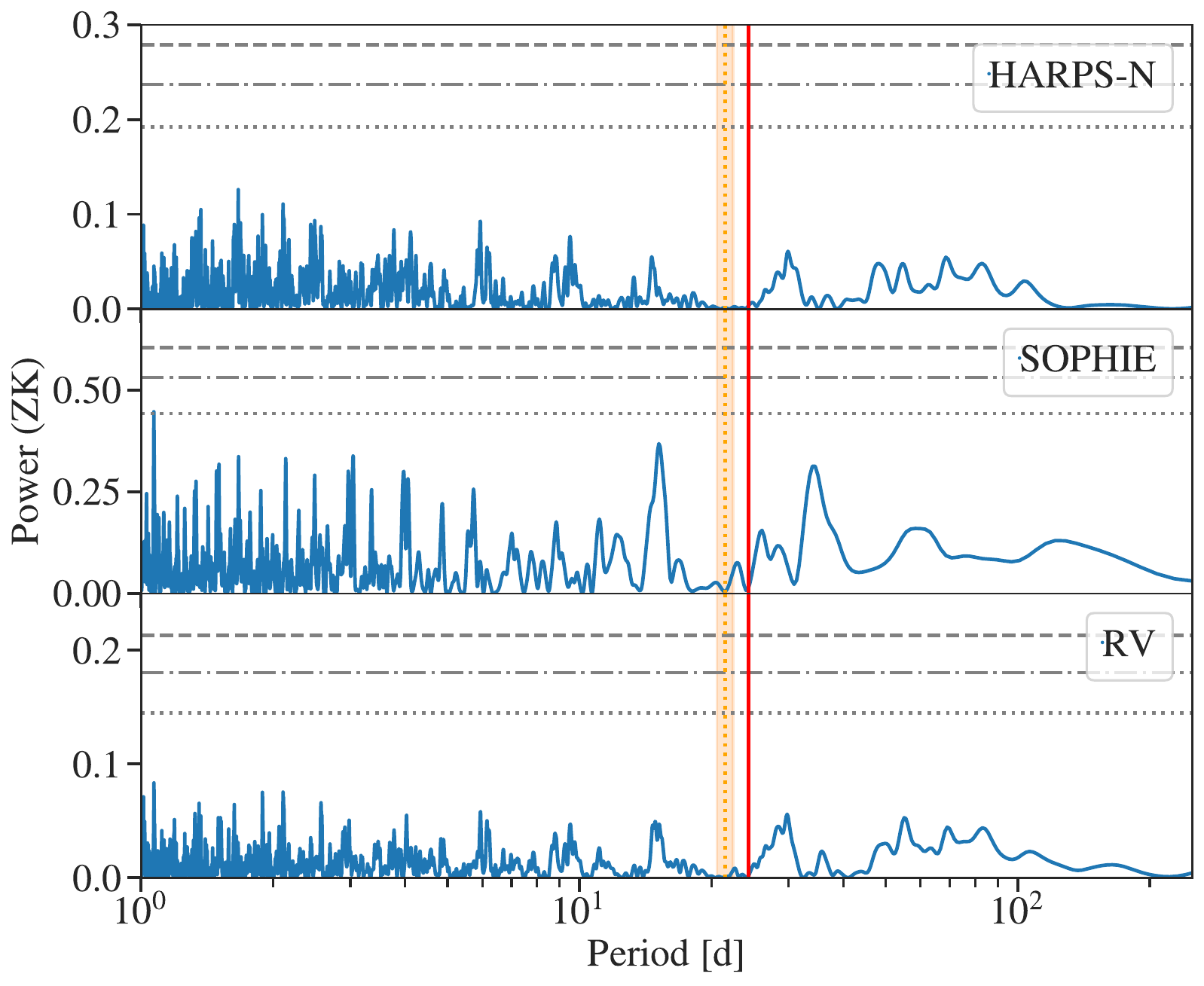}
  \caption{Generalized Lomb-Scargle  periodograms of HARPS-N, SOPHIE, and combined RV residuals (from top to bottom) after subtracting the 1pl model (left), 1pGP model (centre), and 1pGP+FWHM model (right). The vertical red dotted line indicates the period of TOI-1710\,b at 24.28\,d. The shaded orange region indicates the 3$\sigma$ region of the $P_{\rm rot}$ from the joint fit (see Sect.\,\ref{subsec: JF model}).
  The 10\%, 1\%, and 0.1\% FAP levels are indicated by the grey dotted, dash-dotted, and  dashed lines, respectively.
  \label{fig: GLS residuals} }
\end{figure*}


\end{appendix}

\end{document}

%% file: Table_RVs_1.tex
\begin{table*} 
\begin{tiny} 
\begin{center} 
\caption{Radial velocities and spectral activity indicators measured from TNG/HARPS-N spectra with {\tt serval} and DRS. Time stamps from \citet{Konig_TOI1710} spectra are marked in bold. Time series of spectra continue into Table\,\ref{table: table_RV_2}. 
 \label{table: table_RV_1}} 
\begin{tabular}{r rr rr c c rr rr rr} 
\hline \hline \noalign{\smallskip} 
 \multicolumn{1}{c}{BJD$_\mathrm{TBD}$} & \multicolumn{2}{c}{RV \texttt{serval} } & \multicolumn{2}{c}{BIS} & \multicolumn{1}{c}{CCF\_FWHM} & \multicolumn{1}{c}{CCF\_CTR} & \multicolumn{2}{c}{$\mathrm{\log{R^{`}_{HK}}}$} & \multicolumn{2}{c}{CRX} & \multicolumn{2}{c}{dlW} \\  
 \multicolumn{1}{c}{$-$2457000} & \multicolumn{2}{c}{[$\mathrm{m\,s^{-1}}$]} & \multicolumn{2}{c}{[$\mathrm{m\,s^{-1}}$]} & \multicolumn{1}{c}{[$\mathrm{km\,s^{-1}}$]} & \multicolumn{1}{c}{(\%)} & \multicolumn{2}{c}{---} & \multicolumn{2}{c}{($\mathrm{m\,s^{-1}\,Np^{-1}}$)} & \multicolumn{2}{c}{[$\mathrm{m^2\,s^{-2}}$]} \\ 
\multicolumn{1}{c}{Val.} & \multicolumn{1}{c}{Val.} & \multicolumn{1}{c}{$\sigma$} & \multicolumn{1}{c}{Val.} & \multicolumn{1}{c}{$\sigma$} & \multicolumn{1}{c}{Val.} & \multicolumn{1}{c}{Val.} &\multicolumn{1}{c}{Val.} & \multicolumn{1}{c}{$\sigma$} & \multicolumn{1}{c}{Val.} & \multicolumn{1}{c}{$\sigma$} & \multicolumn{1}{c}{Val.} & \multicolumn{1}{c}{$\sigma$} \\ 
\hline \noalign{\smallskip} 
\textbf{2125.664739} & 10.7 & 1.84 & -12.99 & 2.559 &  7.277 & 48.411 & -4.7656 & 0.0105 & 4.747 & 14.819 & 15.794 & 3.639\\ 
\textbf{2126.618253} & 5.06 & 1.67 & -11.391 & 2.08 &  7.266 & 48.511 & -4.7681 & 0.0084 & 2.654 & 13.866 & 7.393 & 2.634\\ 
\textbf{2127.719329} & 2.24 & 1.28 & -10.508 & 1.711 &  7.267 & 48.597 & -4.7574 & 0.0122 & 6.085 & 10.514 & -8.641 & 2.77\\ 
\textbf{2130.73596} & -5.98 & 2.54 & -22.531 & 4.878 &  7.26 & 48.304 & -4.8279 & 0.0149 & 5.319 & 20.869 & 18.585 & 6.192\\ 
\textbf{2133.711696} & -7.68 & 1.35 & -15.256 & 2.305 &  7.254 & 48.632 & -4.7963 & 0.0088 & 8.868 & 10.824 & -11.958 & 3.119\\ 
\textbf{2134.756264} & -8.96 & 1.86 & -18.148 & 2.451 &  7.257 & 48.622 & -4.7663 & 0.0117 & 3.603 & 15.075 & -10.429 & 3.526\\ 
2149.682376 & 5.25 & 1.5 & -11.46 & 2.096 &  7.24 & 48.723 & -4.785 & 0.0071 & -7.997 & 12.821 & -4.596 & 2.747\\ 
2150.688883 & 9.76 & 1.2 & -11.969 & 1.788 &  7.244 & 48.698 & -4.7903 & 0.0061 & -16.119 & 9.969 & -0.516 & 1.78\\ 
2151.723635 & 10.56 & 1.58 & -11.029 & 2.38 &  7.246 & 48.671 & -4.7727 & 0.0097 & 0.421 & 12.7 & 1.356 & 3.282\\ 
\textbf{2156.742015} & -6.31 & 1.12 & -19.71 & 1.724 &  7.266 & 48.596 & -4.7887 & 0.0073 & -7.344 & 9.163 & -5.44 & 2.048\\ 
2158.686506 & -5.64 & 1.54 & -12.062 & 2.384 &  7.221 & 48.765 & -4.7668 & 0.0326 & -7.901 & 12.921 & -12.779 & 3.158\\ 
2160.659656 & -7.27 & 1.24 & -17.6 & 1.771 &  7.226 & 48.806 & -4.7938 & 0.0056 & 9.02 & 10.47 & -13.702 & 2.157\\ 
\textbf{2172.605447} & 9.48 & 1.65 & -11.241 & 2.006 &  7.267 & 48.546 & -4.794 & 0.0054 & 1.762 & 13.263 & 0.551 & 3.3\\ 
\textbf{2188.596847} & 3.8 & 1.3 & -14.871 & 1.96 &  7.272 & 48.48 & -4.7884 & 0.0151 & 10.722 & 10.837 & 11.812 & 2.93\\ 
\textbf{2189.64318} & -2.69 & 1.02 & -12.606 & 1.601 &  7.273 & 48.569 & -4.7596 & 0.0173 & 5.66 & 8.758 & 0.28 & 1.761\\ 
\textbf{2190.610672} & -1.96 & 1.41 & -15.521 & 1.949 &  7.262 & 48.616 & -4.7659 & 0.0222 & 3.008 & 12.432 & -5.422 & 2.553\\ 
\textbf{2191.782237} & 1.06 & 1.07 & -17.256 & 1.394 &  7.264 & 48.64 & -4.771 & 0.0185 & 6.011 & 8.591 & -11.408 & 1.747\\ 
2204.493933 & -7.09 & 1.35 & -20.915 & 2.192 &  7.229 & 48.741 & -4.7981 & 0.0049 & 5.84 & 11.869 & -2.912 & 2.467\\ 
2204.673882 & -5.85 & 1.12 & -20.256 & 1.545 &  7.223 & 48.788 & -4.7929 & 0.0047 & 9.047 & 9.829 & -11.423 & 1.735\\ 
2205.499679 & -1.43 & 1.02 & -14.768 & 1.408 &  7.224 & 48.78 & -4.7944 & 0.0066 & 13.736 & 8.672 & -8.102 & 2.041\\ 
2205.685433 & -2.5 & 1.17 & -10.185 & 1.919 &  7.226 & 48.75 & -4.7814 & 0.0096 & 6.871 & 9.888 & -8.062 & 2.571\\ 
2206.4971 & -0.27 & 1.08 & -19.432 & 1.585 &  7.223 & 48.743 & -4.8021 & 0.0079 & -9.944 & 9.4 & -4.913 & 2.244\\ 
2206.679276 & -4.27 & 2.53 & -8.329 & 4.476 &  7.236 & 48.531 & -4.8129 & 0.0078 & -29.964 & 20.502 & 3.099 & 4.807\\ 
\textbf{2212.534392} & 1.0 & 1.29 & -16.177 & 2.127 &  7.271 & 48.518 & -4.824 & 0.0309 & -11.427 & 10.351 & 3.176 & 3.029\\ 
\textbf{2215.622998} & 15.87 & 1.01 & -20.647 & 1.54 &  7.273 & 48.514 & -4.7824 & 0.0079 & -3.803 & 8.19 & 6.178 & 2.435\\ 
2248.433766 & -1.07 & 1.01 & -27.121 & 1.326 &  7.217 & 48.844 & -4.82 & 0.0142 & -14.392 & 8.621 & -18.587 & 1.964\\ 
2248.444309 & -0.28 & 1.06 & -23.395 & 1.285 &  7.219 & 48.846 & -4.8332 & 0.0133 & 4.076 & 9.451 & -19.282 & 1.514\\ 
2265.415755 & 2.88 & 1.93 & -1.549 & 2.569 &  7.231 & 48.685 & -4.826 & 0.0123 & 21.158 & 15.537 & -1.462 & 3.367\\ 
2265.42869 & 2.53 & 1.63 & -21.879 & 2.93 &  7.229 & 48.657 & -4.8417 & 0.033 & 6.113 & 13.459 & -2.513 & 3.663\\ 
2265.560543 & 2.91 & 2.07 & -6.222 & 3.436 &  7.234 & 48.634 & -4.8317 & 0.0337 & -1.784 & 16.663 & -1.651 & 4.335\\ 
2265.57402 & 3.21 & 2.01 & -7.527 & 3.0 &  7.236 & 48.693 & -4.8179 & 0.0259 & -5.787 & 15.995 & -3.218 & 4.034\\ 
2268.426272 & -0.98 & 0.77 & -16.31 & 1.206 &  7.227 & 48.803 & -4.8096 & 0.0119 & -10.589 & 6.668 & -12.481 & 1.832\\ 
2268.438999 & -0.13 & 0.83 & -14.652 & 1.18 &  7.231 & 48.795 & -4.8162 & 0.0118 & -11.144 & 7.11 & -13.971 & 1.68\\ 
2268.505815 & 0.17 & 1.14 & -12.733 & 1.455 &  7.227 & 48.799 & -4.8228 & 0.0172 & 2.831 & 9.839 & -14.944 & 1.93\\ 
2268.519837 & -0.18 & 1.06 & -14.503 & 1.857 &  7.232 & 48.759 & -4.8353 & 0.0102 & 0.231 & 8.992 & -9.382 & 2.118\\ 
\textbf{2275.338169} & 4.42 & 0.93 & -9.129 & 1.361 &  7.282 & 48.475 & -4.8152 & 0.0121 & -16.989 & 7.845 & 42.678 & 2.134\\ 
\textbf{2276.48642} & -1.04 & 1.25 & -11.031 & 1.735 &  7.269 & 48.524 & -4.7992 & 0.0107 & -23.013 & 10.447 & 5.285 & 2.225\\ 
\textbf{2277.409061} & -0.89 & 1.15 & -11.649 & 1.486 &  7.265 & 48.585 & -4.8085 & 0.0118 & -26.776 & 9.186 & -2.286 & 2.006\\ 
\textbf{2278.43132} & -3.63 & 1.33 & -14.517 & 1.629 &  7.268 & 48.589 & -4.7943 & 0.0117 & -19.579 & 11.132 & -3.436 & 1.95\\ 
\textbf{2287.3862} & 0.35 & 1.44 & -18.127 & 2.659 &  7.248 & 48.629 & -4.8129 & 0.0103 & -6.68 & 11.78 & -5.117 & 3.686\\ 
\textbf{2289.40115} & -3.2 & 1.31 & -16.69 & 1.773 &  7.255 & 48.632 & -4.8181 & 0.0101 & -15.986 & 10.631 & -11.952 & 2.457\\ 
\textbf{2290.341493} & 1.41 & 1.08 & -20.809 & 1.347 &  7.259 & 48.672 & -4.8113 & 0.0132 & -13.846 & 8.778 & 19.952 & 1.925\\ 
\textbf{2294.435475} & 11.65 & 1.34 & -19.321 & 1.759 &  7.282 & 48.512 & -4.7985 & 0.0179 & -21.659 & 10.892 & 3.553 & 2.2\\ 
\textbf{2297.388529} & 4.62 & 1.11 & -12.736 & 1.367 &  7.277 & 48.519 & -4.7866 & 0.0161 & -21.198 & 8.954 & 6.262 & 2.071\\ 
\textbf{2298.383859} & 7.51 & 1.46 & -13.666 & 1.986 &  7.272 & 48.512 & -4.847 & 0.0223 & -29.481 & 11.714 & 5.659 & 2.732\\ 
\textbf{2299.452091} & 1.09 & 1.98 & -16.992 & 3.415 &  7.275 & 48.391 & -4.8706 & 0.0307 & -11.599 & 16.296 & 17.651 & 4.4\\ 
\textbf{2303.371341} & -0.39 & 1.37 & -16.247 & 2.125 &  7.261 & 48.583 & -4.8367 & 0.0441 & -6.505 & 10.725 & -7.953 & 3.33\\ 
\textbf{2304.382026} & -7.29 & 1.8 & -25.52 & 2.417 &  7.266 & 48.606 & -4.815 & 0.0165 & -16.113 & 14.563 & -6.493 & 3.38\\ 
\textbf{2307.349721} & -2.59 & 1.15 & -18.116 & 1.685 &  7.256 & 48.595 & -4.8167 & 0.0182 & -8.408 & 9.098 & 24.798 & 2.865\\ 
2309.358376 & -0.15 & 1.3 & -16.503 & 1.663 &  7.219 & 48.808 & -4.8179 & 0.0091 & 2.781 & 10.239 & -12.948 & 2.071\\ 
2309.371202 & -4.7 & 1.29 & -14.803 & 1.615 &  7.221 & 48.788 & -4.8248 & 0.0104 & -3.381 & 10.303 & -16.168 & 2.449\\ 
2310.348335 & 2.24 & 2.7 & -17.647 & 4.171 &  7.218 & 48.549 & -4.8028 & 0.01 & -21.734 & 20.574 & 38.414 & 5.807\\ 
2310.388924 & -0.26 & 1.01 & -18.428 & 1.663 &  7.226 & 48.777 & -4.8568 & 0.0167 & -18.036 & 7.356 & -11.074 & 2.285\\ 
\textbf{2322.354105} & -4.91 & 2.04 & -6.607 & 2.953 &  7.252 & 48.389 & -4.8559 & 0.0308 & 10.087 & 17.174 & 44.35 & 3.227\\ 
\textbf{2323.352913} & -1.43 & 1.26 & -20.516 & 1.744 &  7.258 & 48.563 & -4.7567 & 0.0136 & 2.532 & 10.796 & 27.362 & 2.544\\ 
\textbf{2324.363167} & -1.08 & 1.52 & -14.854 & 2.11 &  7.247 & 48.657 & -4.7616 & 0.0102 & 7.458 & 12.643 & -20.307 & 3.142\\ 
2469.724254 & -9.49 & 1.48 & -27.555 & 2.283 &  7.219 & 48.807 & -4.7641 & 0.0077 & 21.531 & 12.308 & -17.313 & 2.829\\ 
2469.735454 & -11.03 & 1.32 & -16.553 & 2.147 &  7.22 & 48.824 & -4.8477 & 0.0423 & -3.082 & 11.466 & -19.027 & 3.0\\ 
2469.745825 & -9.52 & 1.4 & -17.314 & 2.067 &  7.221 & 48.815 & -4.8086 & 0.0134 & 26.204 & 11.819 & -14.525 & 2.95\\ 
2470.700788 & -5.58 & 2.52 & -24.563 & 3.877 &  7.22 & 48.715 & -4.808 & 0.0151 & 15.515 & 20.619 & -17.841 & 4.693\\ 
2470.722284 & -7.18 & 2.48 & -12.336 & 4.005 &  7.224 & 48.735 & -4.7924 & 0.0085 & -22.477 & 20.357 & -18.624 & 5.05\\ 
2470.741717 & -10.08 & 2.04 & -14.927 & 3.423 &  7.213 & 48.738 & -4.7746 & 0.0102 & -17.234 & 16.714 & -9.654 & 3.832\\ 
2471.706179 & -4.7 & 1.49 & -24.021 & 2.049 &  7.215 & 48.854 & -4.7648 & 0.0101 & -1.31 & 12.23 & -21.572 & 2.816\\ 
2471.71689 & -4.41 & 1.34 & -19.861 & 2.018 &  7.218 & 48.829 & -4.7567 & 0.0072 & 5.007 & 11.085 & -17.867 & 2.701\\ 
2548.691205 & -2.68 & 1.2 & -20.396 & 2.187 &  7.201 & 48.946 & -4.7777 & 0.0104 & 8.304 & 10.706 & -32.646 & 2.794\\ 
2548.715527 & -2.12 & 1.05 & -23.792 & 1.561 &  7.209 & 48.93 & -4.7743 & 0.0061 & -14.167 & 9.375 & -28.177 & 2.083\\ 
2549.68441 & -2.72 & 1.39 & -19.188 & 1.844 &  7.21 & 48.891 & -4.7593 & 0.0109 & 6.16 & 12.284 & -23.636 & 2.341\\ 
2549.694997 & -2.85 & 1.28 & -17.059 & 1.745 &  7.209 & 48.896 & -4.7545 & 0.0065 & 12.199 & 11.302 & -20.913 & 2.058\\ 
2571.606297 & -10.24 & 1.31 & -21.194 & 1.797 &  7.205 & 48.898 & -4.7508 & 0.0055 & 8.551 & 12.454 & -25.188 & 2.151\\ 
2571.617012 & -8.73 & 1.46 & -20.827 & 1.819 &  7.202 & 48.9 & -4.7545 & 0.0081 & 9.234 & 13.847 & -25.299 & 2.076\\ 
\noalign{\smallskip} \hline 
\end{tabular} 
\end{center} 
\end{tiny} 
\end{table*} 

%% file: Table_ACT_1.tex
\begin{table*} 
\begin{tiny} 
\begin{center} 
\caption{Continuation of radial velocities and spectral activity indicators from Table\,\ref{table: table_RV_1}. 
 \label{table: table_ACT_1}} 
\begin{tabular}{r rr rr rr rr r} 
\hline \hline \noalign{\smallskip} 
\multicolumn{1}{c}{BJD$_\mathrm{TBD}$} & \multicolumn{2}{c}{DRS RV} & \multicolumn{2}{c}{$\mathrm{H\alpha}$} & \multicolumn{2}{c}{$\mathrm{NaD_{1}}$} & \multicolumn{2}{c}{$\mathrm{NaD_{2}}$} & \multicolumn{1}{c}{SNR}  \\ 
 \multicolumn{1}{c}{$-$2457000} & \multicolumn{2}{c}{[$\mathrm{m\,s^{-1}}$]} & \multicolumn{2}{c}{---} & \multicolumn{2}{c}{---} & \multicolumn{2}{c}{---} & \multicolumn{1}{c}{(@550nm)} \\ 
\hline \noalign{\smallskip} 
\multicolumn{1}{c}{Val.} & \multicolumn{1}{c}{Val.} & \multicolumn{1}{c}{$\sigma$} &  \multicolumn{1}{c}{Val.} & \multicolumn{1}{c}{$\sigma$} & \multicolumn{1}{c}{Val.} & \multicolumn{1}{c}{$\sigma$} & \multicolumn{1}{c}{Val.} & \multicolumn{1}{c}{$\sigma$} &  \multicolumn{1}{c}{Val.} \\ 
\textbf{2125.664739} & -38807.06 & 1.79 & 0.4358 & 0.0014 & 0.3151 & 0.0017 & 0.4121 & 0.0021 & 50.4\\ 
\textbf{2126.618253} & -38813.01 & 1.46 & 0.4375 & 0.0012 & 0.3169 & 0.0015 & 0.4211 & 0.0018 & 61.3\\ 
\textbf{2127.719329} & -38815.83 & 1.2 & 0.4381 & 0.001 & 0.3201 & 0.0012 & 0.4195 & 0.0015 & 71.9\\ 
\textbf{2130.73596} & -38826.8 & 3.42 & 0.4356 & 0.0026 & 0.3195 & 0.0032 & 0.4395 & 0.0039 & 30.2\\ 
\textbf{2133.711696} & -38826.62 & 1.62 & 0.4303 & 0.0012 & 0.3105 & 0.0016 & 0.4196 & 0.002 & 54.8\\ 
\textbf{2134.756264} & -38828.23 & 1.72 & 0.4326 & 0.0013 & 0.3084 & 0.0017 & 0.4236 & 0.0021 & 52.3\\ 
2149.682376 & -38812.1 & 1.48 & 0.4368 & 0.0014 & 0.3134 & 0.0015 & 0.429 & 0.0019 & 61.7\\ 
2150.688883 & -38804.86 & 1.26 & 0.4361 & 0.0011 & 0.3172 & 0.0013 & 0.4168 & 0.0016 & 71.4\\ 
2151.723635 & -38808.6 & 1.68 & 0.4358 & 0.0012 & 0.3072 & 0.0016 & 0.4167 & 0.002 & 54.7\\ 
\textbf{2156.742015} & -38823.64 & 1.21 & 0.4337 & 0.001 & 0.3135 & 0.0012 & 0.4215 & 0.0015 & 71.8\\ 
2158.686506 & -38823.02 & 1.68 & 0.429 & 0.0014 & 0.3199 & 0.0016 & 0.4174 & 0.002 & 55.3\\ 
2160.659656 & -38825.07 & 1.25 & 0.4284 & 0.0011 & 0.3132 & 0.0012 & 0.4202 & 0.0016 & 71.8\\ 
\textbf{2172.605447} & -38807.53 & 1.4 & 0.4345 & 0.0011 & 0.3162 & 0.0014 & 0.4183 & 0.0017 & 62.4\\ 
\textbf{2188.596847} & -38814.75 & 1.38 & 0.4329 & 0.0012 & 0.3177 & 0.0014 & 0.4216 & 0.0018 & 63.9\\ 
\textbf{2189.64318} & -38821.64 & 1.12 & 0.4312 & 0.0011 & 0.3129 & 0.0012 & 0.4161 & 0.0015 & 76.8\\ 
\textbf{2190.610672} & -38820.25 & 1.37 & 0.4303 & 0.0013 & 0.3171 & 0.0015 & 0.415 & 0.0018 & 63.8\\ 
\textbf{2191.782237} & -38819.01 & 0.98 & 0.4332 & 0.0007 & 0.3182 & 0.001 & 0.4151 & 0.0012 & 87.7\\ 
2204.493933 & -38824.26 & 1.55 & 0.4318 & 0.0015 & 0.3187 & 0.0016 & 0.4142 & 0.002 & 59.3\\ 
2204.673882 & -38821.82 & 1.09 & 0.4338 & 0.0011 & 0.3107 & 0.0012 & 0.4277 & 0.0014 & 80.9\\ 
2205.499679 & -38817.96 & 0.99 & 0.4315 & 0.0009 & 0.3181 & 0.001 & 0.4193 & 0.0013 & 88.0\\ 
2205.685433 & -38819.52 & 1.35 & 0.433 & 0.0012 & 0.3167 & 0.0014 & 0.4157 & 0.0017 & 66.7\\ 
2206.4971 & -38818.84 & 1.12 & 0.4339 & 0.0011 & 0.3187 & 0.0012 & 0.4172 & 0.0015 & 79.2\\ 
2206.679276 & -38823.34 & 3.16 & 0.4351 & 0.0022 & 0.3229 & 0.0029 & 0.4216 & 0.0035 & 33.0\\ 
\textbf{2212.534392} & -38817.57 & 1.49 & 0.4333 & 0.0011 & 0.3178 & 0.0015 & 0.4185 & 0.0018 & 58.8\\ 
\textbf{2215.622998} & -38804.38 & 1.08 & 0.4338 & 0.0008 & 0.3178 & 0.0011 & 0.4176 & 0.0013 & 78.7\\ 
2248.433766 & -38817.36 & 0.94 & 0.4353 & 0.0009 & 0.3119 & 0.001 & 0.4201 & 0.0012 & 93.9\\ 
2248.444309 & -38817.29 & 0.91 & 0.4365 & 0.001 & 0.3103 & 0.001 & 0.4197 & 0.0012 & 96.6\\ 
2265.415755 & -38812.63 & 1.81 & 0.4353 & 0.0014 & 0.3143 & 0.0017 & 0.419 & 0.0022 & 51.8\\ 
2265.42869 & -38812.7 & 2.06 & 0.4364 & 0.0016 & 0.315 & 0.0019 & 0.4193 & 0.0024 & 46.7\\ 
2265.560543 & -38814.58 & 2.42 & 0.4368 & 0.0016 & 0.3094 & 0.0022 & 0.4159 & 0.0027 & 40.6\\ 
2265.57402 & -38812.17 & 2.11 & 0.435 & 0.0014 & 0.3144 & 0.0019 & 0.4202 & 0.0024 & 45.3\\ 
2268.426272 & -38816.12 & 0.85 & 0.4335 & 0.0008 & 0.3069 & 0.0009 & 0.4173 & 0.0011 & 103.8\\ 
2268.438999 & -38816.24 & 0.83 & 0.4351 & 0.0008 & 0.3076 & 0.0009 & 0.4187 & 0.0011 & 106.2\\ 
2268.505815 & -38816.0 & 1.03 & 0.4345 & 0.0009 & 0.3132 & 0.001 & 0.4181 & 0.0013 & 87.5\\ 
2268.519837 & -38815.97 & 1.31 & 0.4337 & 0.0011 & 0.314 & 0.0013 & 0.4197 & 0.0016 & 70.1\\ 
\textbf{2275.338169} & -38812.94 & 0.95 & 0.4331 & 0.0009 & 0.3135 & 0.001 & 0.4218 & 0.0013 & 90.9\\ 
\textbf{2276.48642} & -38820.71 & 1.22 & 0.4378 & 0.0012 & 0.3111 & 0.0013 & 0.42 & 0.0016 & 73.3\\ 
\textbf{2277.409061} & -38820.58 & 1.04 & 0.4397 & 0.001 & 0.3129 & 0.0011 & 0.4224 & 0.0014 & 83.7\\ 
\textbf{2278.43132} & -38822.2 & 1.14 & 0.4363 & 0.0011 & 0.3144 & 0.0012 & 0.4228 & 0.0015 & 77.4\\ 
\textbf{2287.3862} & -38816.83 & 1.86 & 0.4379 & 0.0015 & 0.3131 & 0.0018 & 0.4257 & 0.0023 & 49.6\\ 
\textbf{2289.40115} & -38820.9 & 1.24 & 0.4403 & 0.001 & 0.3121 & 0.0013 & 0.4202 & 0.0016 & 70.6\\ 
\textbf{2290.341493} & -38816.6 & 0.94 & 0.4396 & 0.0008 & 0.3131 & 0.001 & 0.4215 & 0.0012 & 91.0\\ 
\textbf{2294.435475} & -38806.25 & 1.23 & 0.4341 & 0.0011 & 0.3132 & 0.0012 & 0.42 & 0.0016 & 72.5\\ 
\textbf{2297.388529} & -38813.86 & 0.96 & 0.4364 & 0.0009 & 0.3136 & 0.001 & 0.4224 & 0.0012 & 90.9\\ 
\textbf{2298.383859} & -38810.32 & 1.39 & 0.4384 & 0.0012 & 0.3126 & 0.0014 & 0.4199 & 0.0018 & 65.1\\ 
\textbf{2299.452091} & -38816.21 & 2.39 & 0.437 & 0.0018 & 0.311 & 0.0022 & 0.4233 & 0.0028 & 41.2\\ 
\textbf{2303.371341} & -38819.51 & 1.49 & 0.4382 & 0.001 & 0.3153 & 0.0014 & 0.4228 & 0.0018 & 59.3\\ 
\textbf{2304.382026} & -38824.48 & 1.69 & 0.4379 & 0.0013 & 0.3115 & 0.0016 & 0.4191 & 0.0021 & 54.1\\ 
\textbf{2307.349721} & -38821.72 & 1.18 & 0.433 & 0.0009 & 0.3125 & 0.0011 & 0.4312 & 0.0014 & 73.5\\ 
2309.358376 & -38817.17 & 1.17 & 0.4319 & 0.0008 & 0.3092 & 0.0011 & 0.4224 & 0.0014 & 75.4\\ 
2309.371202 & -38820.39 & 1.14 & 0.433 & 0.0008 & 0.3082 & 0.0011 & 0.4223 & 0.0014 & 78.3\\ 
2310.348335 & -38814.24 & 2.94 & 0.4361 & 0.0017 & 0.3121 & 0.0025 & 0.4206 & 0.0031 & 34.2\\ 
2310.388924 & -38815.96 & 1.17 & 0.4368 & 0.0007 & 0.3121 & 0.0011 & 0.4248 & 0.0014 & 73.3\\ 
\textbf{2322.354105} & -38822.48 & 2.07 & 0.4274 & 0.0016 & 0.3151 & 0.002 & 0.4295 & 0.0025 & 45.8\\ 
\textbf{2323.352913} & -38818.13 & 1.22 & 0.432 & 0.0011 & 0.3104 & 0.0013 & 0.4193 & 0.0016 & 72.2\\ 
\textbf{2324.363167} & -38821.12 & 1.48 & 0.431 & 0.0012 & 0.308 & 0.0015 & 0.4138 & 0.0018 & 61.0\\ 
2469.724254 & -38826.37 & 1.61 & 0.4368 & 0.0014 & 0.3161 & 0.0016 & 0.4203 & 0.002 & 57.9\\ 
2469.735454 & -38827.19 & 1.51 & 0.4364 & 0.0014 & 0.313 & 0.0015 & 0.4183 & 0.0019 & 61.0\\ 
2469.745825 & -38825.9 & 1.46 & 0.4328 & 0.0014 & 0.317 & 0.0015 & 0.4153 & 0.0019 & 63.0\\ 
2470.700788 & -38823.39 & 2.73 & 0.4331 & 0.002 & 0.3171 & 0.0025 & 0.4179 & 0.0031 & 37.1\\ 
2470.722284 & -38821.79 & 2.82 & 0.434 & 0.0021 & 0.316 & 0.0026 & 0.4202 & 0.0032 & 36.1\\ 
2470.741717 & -38826.41 & 2.41 & 0.4352 & 0.0018 & 0.3164 & 0.0023 & 0.4137 & 0.0028 & 41.0\\ 
2471.706179 & -38822.72 & 1.44 & 0.4295 & 0.0011 & 0.3179 & 0.0014 & 0.4189 & 0.0017 & 63.4\\ 
2471.71689 & -38821.18 & 1.42 & 0.4323 & 0.0012 & 0.3188 & 0.0014 & 0.4177 & 0.0017 & 64.6\\ 
2548.691205 & -38819.07 & 1.54 & 0.4306 & 0.0013 & 0.3183 & 0.0015 & 0.4301 & 0.0018 & 62.6\\ 
2548.715527 & -38818.63 & 1.1 & 0.4282 & 0.0009 & 0.3166 & 0.0011 & 0.4326 & 0.0013 & 84.5\\ 
2549.68441 & -38819.35 & 1.3 & 0.429 & 0.0011 & 0.3174 & 0.0012 & 0.4241 & 0.0015 & 72.9\\ 
2549.694997 & -38821.02 & 1.23 & 0.4304 & 0.001 & 0.3183 & 0.0012 & 0.4217 & 0.0015 & 76.9\\ 
2571.606297 & -38825.54 & 1.27 & 0.4273 & 0.0012 & 0.3173 & 0.0012 & 0.4239 & 0.0016 & 75.2\\ 
2571.617012 & -38825.13 & 1.28 & 0.4276 & 0.0012 & 0.3176 & 0.0013 & 0.4259 & 0.0016 & 74.5\\ 
\noalign{\smallskip} \hline 
\end{tabular} 
\end{center} 
\end{tiny} 
\end{table*}

%% file: Table_RVs_2.tex
\begin{table*} 
\begin{tiny} 
\begin{center} 
\caption{More radial velocities and spectral activity indicators measured from TNG/HARPS-N spectra with {\tt serval} and DRS, continuation from Table\,\ref{table: table_RV_1}. 
 \label{table: table_RV_2}} 
\begin{tabular}{r rr rr c c rr rr rr} 
\hline \hline \noalign{\smallskip} 
 \multicolumn{1}{c}{BJD$_\mathrm{TBD}$} & \multicolumn{2}{c}{RV \texttt{serval} } & \multicolumn{2}{c}{BIS} & \multicolumn{1}{c}{CCF\_FWHM} & \multicolumn{1}{c}{CCF\_CTR} & \multicolumn{2}{c}{$\mathrm{\log{R^{`}_{HK}}}$} & \multicolumn{2}{c}{CRX} & \multicolumn{2}{c}{dlW} \\  
 \multicolumn{1}{c}{$-$2457000} & \multicolumn{2}{c}{[$\mathrm{m\,s^{-1}}$]} & \multicolumn{2}{c}{[$\mathrm{m\,s^{-1}}$]} & \multicolumn{1}{c}{[$\mathrm{km\,s^{-1}}$]} & \multicolumn{1}{c}{(\%)} & \multicolumn{2}{c}{---} & \multicolumn{2}{c}{($\mathrm{m\,s^{-1}\,Np^{-1}}$)} & \multicolumn{2}{c}{[$\mathrm{m^2\,s^{-2}}$]} \\ 
\multicolumn{1}{c}{Val.} & \multicolumn{1}{c}{Val.} & \multicolumn{1}{c}{$\sigma$} & \multicolumn{1}{c}{Val.} & \multicolumn{1}{c}{$\sigma$} & \multicolumn{1}{c}{Val.} & \multicolumn{1}{c}{Val.} &\multicolumn{1}{c}{Val.} & \multicolumn{1}{c}{$\sigma$} & \multicolumn{1}{c}{Val.} & \multicolumn{1}{c}{$\sigma$} & \multicolumn{1}{c}{Val.} & \multicolumn{1}{c}{$\sigma$} \\ 
\hline \noalign{\smallskip} 
2572.618783 & -9.92 & 1.1 & -16.587 & 1.654 &  7.203 & 48.926 & -4.7687 & 0.0063 & 0.712 & 10.598 & -26.164 & 2.053\\ 
2573.609105 & -8.63 & 1.05 & -16.946 & 1.609 &  7.205 & 48.935 & -4.7727 & 0.0073 & -2.969 & 10.216 & -30.159 & 2.145\\ 
2573.620272 & -10.03 & 1.42 & -19.354 & 1.91 &  7.196 & 48.96 & -4.7861 & 0.0167 & -4.553 & 13.573 & -35.489 & 2.397\\ 
2574.616426 & -9.14 & 1.21 & -24.856 & 1.892 &  7.206 & 48.922 & -4.7958 & 0.009 & 2.954 & 11.396 & -32.579 & 2.343\\ 
2574.626909 & -5.63 & 1.22 & -20.424 & 1.761 &  7.203 & 48.981 & -4.7911 & 0.0057 & 6.312 & 11.505 & -36.546 & 2.263\\ 
2593.541588 & -12.46 & 1.45 & -21.182 & 2.575 &  7.195 & 48.881 & -4.7512 & 0.0078 & 23.799 & 13.114 & -28.16 & 3.367\\ 
2594.558159 & -13.36 & 2.09 & -25.055 & 3.025 &  7.201 & 48.767 & -4.7486 & 0.0054 & 12.099 & 18.98 & -16.529 & 3.449\\ 
2608.642185 & -2.63 & 2.25 & -16.671 & 4.002 &  7.202 & 48.897 & -4.7303 & 0.0093 & 10.776 & 20.233 & -28.475 & 4.352\\ 
2609.541599 & -6.15 & 1.3 & -12.365 & 2.229 &  7.189 & 48.99 & -4.7497 & 0.0215 & 8.681 & 11.756 & -36.94 & 2.724\\ 
2609.60981 & -2.56 & 1.92 & -14.308 & 2.344 &  7.203 & 48.998 & -4.7685 & 0.0109 & 5.883 & 17.278 & -34.373 & 3.209\\ 
2610.443516 & -4.76 & 1.31 & -16.717 & 1.539 &  7.202 & 49.008 & -4.8084 & 0.0151 & 32.462 & 12.065 & -34.381 & 1.831\\ 
2610.564053 & -9.59 & 1.17 & -16.357 & 1.646 &  7.202 & 49.018 & -4.8002 & 0.0084 & 29.505 & 10.587 & -36.771 & 2.288\\ 
2664.36496 & -8.65 & 1.36 & -15.026 & 1.651 &  7.209 & 48.987 & -4.8028 & 0.021 & 15.094 & 12.691 & -31.371 & 2.305\\ 
2676.357574 & -13.89 & 1.5 & -21.413 & 2.078 &  7.191 & 49.086 & -4.7978 & 0.0091 & 11.69 & 14.028 & -45.617 & 2.586\\ 
2677.355196 & -13.72 & 1.74 & -19.361 & 3.014 &  7.195 & 49.0 & -4.8078 & 0.0122 & -13.236 & 16.024 & -5.524 & 3.676\\ 
\noalign{\smallskip} \hline 
\end{tabular} 
\end{center} 
\end{tiny} 
\end{table*} 

%% file: Table_ACT_2.tex
\begin{table*} 
\begin{tiny} 
\begin{center} 
\caption{Continuation of radial velocities and spectral activity indicators from Table\,\ref{table: table_RV_2}. 
 \label{table: table_ACT_2}} 
\begin{tabular}{r rr rr rr rr r} 
\hline \hline \noalign{\smallskip} 
\multicolumn{1}{c}{BJD$_\mathrm{TBD}$} & \multicolumn{2}{c}{DRS RV} & \multicolumn{2}{c}{$\mathrm{H\alpha}$} & \multicolumn{2}{c}{$\mathrm{NaD_{1}}$} & \multicolumn{2}{c}{$\mathrm{NaD_{2}}$} & \multicolumn{1}{c}{SNR}  \\ 
 \multicolumn{1}{c}{$-$2457000} & \multicolumn{2}{c}{[$\mathrm{m\,s^{-1}}$]} & \multicolumn{2}{c}{---} & \multicolumn{2}{c}{---} & \multicolumn{2}{c}{---} & \multicolumn{1}{c}{(@550nm)} \\ 
\hline \noalign{\smallskip} 
\multicolumn{1}{c}{Val.} & \multicolumn{1}{c}{Val.} & \multicolumn{1}{c}{$\sigma$} &  \multicolumn{1}{c}{Val.} & \multicolumn{1}{c}{$\sigma$} & \multicolumn{1}{c}{Val.} & \multicolumn{1}{c}{$\sigma$} & \multicolumn{1}{c}{Val.} & \multicolumn{1}{c}{$\sigma$} & \multicolumn{1}{c}{Val.} \\ 
2572.618783 & -38827.33 & 1.17 & 0.4286 & 0.0011 & 0.3194 & 0.0012 & 0.424 & 0.0015 & 81.0\\ 
2573.609105 & -38826.06 & 1.13 & 0.4317 & 0.0011 & 0.3146 & 0.0011 & 0.4231 & 0.0014 & 83.2\\ 
2573.620272 & -38826.21 & 1.35 & 0.4295 & 0.0013 & 0.3154 & 0.0014 & 0.4259 & 0.0017 & 71.2\\ 
2574.616426 & -38827.24 & 1.33 & 0.4349 & 0.0011 & 0.3223 & 0.0012 & 0.426 & 0.0015 & 75.0\\ 
2574.626909 & -38824.05 & 1.24 & 0.4309 & 0.001 & 0.319 & 0.0011 & 0.4264 & 0.0014 & 79.8\\ 
2593.541588 & -38829.8 & 1.82 & 0.426 & 0.0016 & 0.3166 & 0.0017 & 0.4228 & 0.0022 & 55.2\\ 
2594.558159 & -38831.89 & 2.13 & 0.4286 & 0.0018 & 0.3154 & 0.002 & 0.4269 & 0.0025 & 48.1\\ 
2608.642185 & -38823.24 & 2.82 & 0.4302 & 0.0022 & 0.3101 & 0.0025 & 0.4237 & 0.0032 & 38.9\\ 
2609.541599 & -38821.59 & 1.57 & 0.4266 & 0.0012 & 0.3101 & 0.0014 & 0.4285 & 0.0018 & 61.8\\ 
2609.60981 & -38818.43 & 1.65 & 0.4257 & 0.0012 & 0.3115 & 0.0015 & 0.4355 & 0.0019 & 59.2\\ 
2610.443516 & -38822.51 & 1.09 & 0.429 & 0.0009 & 0.3127 & 0.001 & 0.4246 & 0.0013 & 86.9\\ 
2610.564053 & -38826.56 & 1.16 & 0.4263 & 0.001 & 0.3145 & 0.0011 & 0.4255 & 0.0014 & 82.0\\ 
2664.36496 & -38826.09 & 1.16 & 0.4215 & 0.001 & 0.3137 & 0.0011 & 0.4346 & 0.0014 & 81.0\\ 
2676.357574 & -38830.48 & 1.46 & 0.4189 & 0.0014 & 0.3138 & 0.0015 & 0.4225 & 0.0019 & 65.9\\ 
2677.355196 & -38829.55 & 2.12 & 0.4224 & 0.0018 & 0.3151 & 0.002 & 0.4357 & 0.0025 & 48.6\\ 
\noalign{\smallskip} \hline 
\end{tabular} 
\end{center} 
\end{tiny} 
\end{table*}